\documentclass[%
 reprint,superscriptaddress,
 amsmath,amssymb,prx]{revtex4-2}

\usepackage{physics}
\usepackage{xcolor}
\usepackage{soul}
\usepackage{xtab,afterpage,longtable}
\usepackage{lipsum}
\usepackage{graphicx}
\usepackage{dcolumn}
\usepackage{booktabs} 
\usepackage{multirow}
\usepackage{bbm}  
\usepackage{color}
\usepackage{csquotes}
\usepackage{hyperref}
\usepackage{soul}

\makeatletter
\renewcommand*{\@fnsymbol}[1]{\ensuremath{\ifcase#1\or \dagger\or *\or \ddagger\or \mathsection\or \mathparagraph\or \|\or **\or \dagger\dagger\or \ddagger\ddagger\else\@ctrerr\fi}}
\makeatother

\begin{document}

\title{Deep Learning of Mean First Passage Time Scape: \\Chemical Short-Range Order and Kinetics of Diffusive Relaxation}

\author{Hoje Chun}
\thanks{These authors contributed equally to this work.}
\affiliation{Department of Materials Science and Engineering, Massachusetts Institute of Technology, Cambridge, MA 02139, USA}

\author{Hao Tang}
\thanks{These authors contributed equally to this work.}
\affiliation{Department of Materials Science and Engineering, Massachusetts Institute of Technology, Cambridge, MA 02139, USA}

\author{Bin Xing}
\affiliation{Department of Nuclear Science and Engineering, Massachusetts Institute of Technology, Cambridge, MA 02139, USA}

\author{Rafael G\'{o}mez-Bombarelli}
\affiliation{Department of Materials Science and Engineering, Massachusetts Institute of Technology, Cambridge, MA 02139, USA}

\author{Ju Li}
\thanks{Corresponding author: liju@mit.edu}
\affiliation{Department of Materials Science and Engineering, Massachusetts Institute of Technology, Cambridge, MA 02139, USA}
\affiliation{Department of Nuclear Science and Engineering, Massachusetts Institute of Technology, Cambridge, MA 02139, USA}

\date{\today}

\begin{abstract}
Processes slow compared to atomic vibrations pose significant challenges in atomistic simulations, particularly for phenomena such as diffusive relaxations and phase transitions, where repeated crossings and the shear number of thermally activated transitions make direct numerical simulations impossible. 
We present a computational framework that captures atomic-scale diffusive relaxation over extended timescales by learning the mean first passage time (MFPT) with a deep neural network. 
The model is trained via a self-consistent recursive formulation based on the Markovian assumption, relying solely on local residence times and transition probabilities between neighboring states.
Furthermore, we leverage deep reinforcement learning (DRL)-accelerated atomistic simulations to expedite the identification of thermodynamic equilibrium and the generation of accurate physical transition probabilities. 
Applied to vacancy-mediated chemical short-range order (SRO) evolution in equiatomic CrCoNi, our method uncovers disorder-to-order transition timescales in quantitative agreement with experimental measurements.
By bridging the gap between simulation and experiment, our approach extends atomistic modeling to previously inaccessible timescales and offers a predictive tool for navigating process–structure–property relationships.
\end{abstract}

\maketitle

\section{Introduction}
Atomistic simulations have long been confronted with challenges to capture physical phenomena on experimentally relevant timescales due to inherent computational limitations~\cite{uberuaga2020computational}. Many critical material processes, such as phase transitions and mechanical deformation, span a broad spectrum of timescales, often reaching minutes or longer, whereas conventional molecular dynamics (MD) simulations are typically limited to nanoseconds or microseconds~\cite{henkelman2018long, voter2007introduction}. 
Due to the necessity of small integration time steps for simulation stability, the extension of MD simulations to longer timescales remains infeasible, even with machine-learning interatomic potentials (ML-IPs)~\cite{batatia2023foundation, deng2023chgnet, takamoto2022towards}.  
The Arrhenius dependence of transition rates further widens the timescale gap, limiting access to rare events in simulation.
To address this limitation, a variety of accelerated dynamics techniques based on transition-state theory (TST) have been developed, including hyperdynamics~\cite{voter1997hyperdynamics,hara2010adaptive}, parallel replica dynamics~\cite{voter1998parallel}, diffusive molecular dynamics~\cite{li2011diffusive,sarkar2012finding}, or generative modeling~\cite{nam2025flow}. Nevertheless, long-timescale simulation of processes consisting of a large number of atomic diffusion events is still a challenge. 

A prominent example of long-timescale behavior is the formation of chemical short-range order (SRO) in multi-principal-component medium- and high-entropy alloys (MEAs/HEAs)~\cite{zhang2020short,ferrari2023simulating, walsh2023reconsidering,AbdelhafizTZWRHZL23}.  
These alloys have attracted considerable interest due to their superior mechanical, thermal, and catalytic properties compared to conventional alloys, which typically comprise one or two principal elements~\cite{gludovatz2014fracture, yao2018carbothermal, wagner2022effects, otto2013influences}.  
Furthermore, the emergence of SRO can significantly affect key materials properties, including mechanical strength and magnetic behavior~\cite{ziehl2023detection, ding2018tunable, walsh2021magnetically}.  
The SRO formation and its associated property change are typically driven by thermal annealing at elevated temperatures.  
Importantly, the resulting SRO is often not in thermodynamic equilibrium but instead reflects the processing history.

Understanding the processing-structure-property relation of MEAs/HEAs is not a trivial task, mainly because the experimental quantification of the SRO features at the nanometer scale is very challenging~\cite{li2023evolution, he2024quantifying}.
In light of these challenges, computational approaches have emerged as powerful alternatives, especially when coupled with recent advances in machine learning for atomistic simulations.
For instance, the Freitas group has successfully identified the thermodynamic equilibrium atomic configurations via ML-IP accelerated Monte Carlo (MC) sampling and has accurately quantified the SRO with local chemical motif~\cite{sheriff2024quantifying,cao2024capturing}.
The Ogata group has demonstrated a machine learning-integrated kMC simulation of vacancy diffusion in CrCoNi to investigate the kinetics of SRO formation and provided an empirical mapping of the time-temperature-SRO relationship~\cite{li2024tunable,du2022chemical,shen2021kinetic}.
Despite these advances, tracing the physical time evolution of SRO remains elusive. 
Conventional MC methods for thermodynamic equilibrium state sampling generally do not account for the physical transition of atomic configurations.
Furthermore, obtaining a complete kMC trajectory toward thermodynamic equilibrium at lower temperatures such as room temperature is nearly unattainable because of the prohibitively long simulation times to overcome repeated crossing of atomic transition events.


In this work, we address the long-timescale challenge and elucidate the processing-structure relation at the atomic level by learning the mean first passage time (MFPT)~\cite{shlesinger2006search}, the time required for an atomic configuration to reach its thermodynamic equilibrium. 
The mapping from atomic structures to the MFPT is visually in analogy to a potential energy landscape in the configuration space, so we abbreviate it as the ``timescape".
To train the surrogate model of a deep neural network (DNN) for predicting MFPT, we propose a recursive loss function based on a Markovian network, and an update scheme of the temporal difference (TD) learning.
We employ an equivariant graph neural network (GNN) to encode not only local atomic environments but also atomic transitions, enabling accurate prediction of the kinetic and thermodynamic parameters of these events. 
In addition, we effectively construct the timescape dataset from mimetic trajectories that capture both the thermodynamic destination with its fluctuations and the transition probabilities by leveraging deep reinforcement learning (DRL)-accelerated atomistic simulations~\cite{tang2024reinforcement}.
We apply this framework to study the vacancy-mediated evolution of SRO in equiatomic CrCoNi.
The MFPT predictions successfully capture the physical time evolution of SRO formation across temperatures and enable quantitative mapping among SRO, time, and temperature. 
Our approach overcomes the timescale limitations of conventional atomistic simulations, expanding their applicability to a wide range of long-timescale transformation phenomena.
Moreover, this offers valuable insights for optimizing processing conditions to achieve a desired SRO, ultimately enabling the design of materials with target properties.

\section{Results}
\subsection{Conceptual framework}\label{conceptual_framework}

\begin{figure*}[!htb]\centering\includegraphics{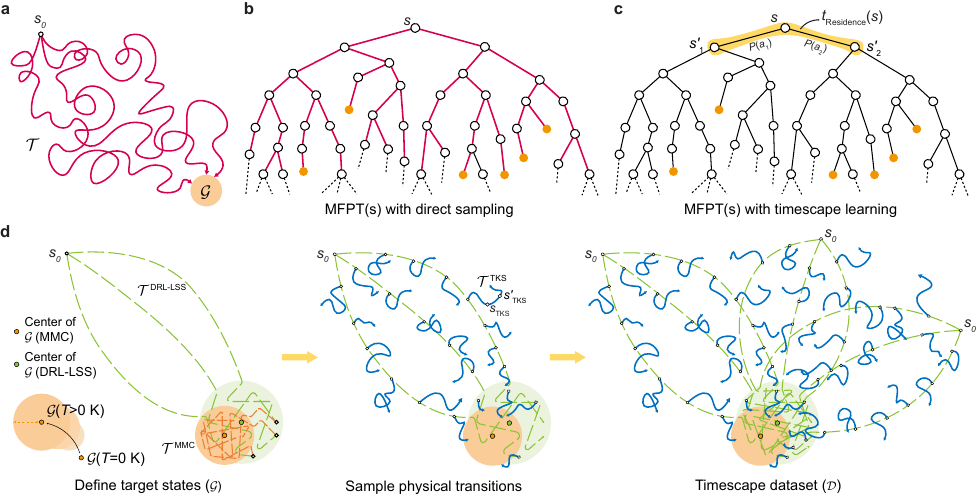}
    \caption{Schematic illustration of diffusive relaxation kinetics and learning of the MFPT timescape using DRL atomistic simulations. 
\textbf{a}, Complete physical trajectories, $\{\mathcal{T}\}$ from a random solid solution, $s_0$, to the target states ($\mathcal{G}$) in a two-dimensional projected configuration space. The orange sphere represents the volume of $\mathcal G$.
\textbf{b}, Evaluation of MFPT($s$) by direct sampling. Circles represent energy minima, and lines denote physical transition pathways. States in $\mathcal{G}$ are highlighted in orange.
\textbf{c}, Timescape learning used in this work. The MFPT($s$) is trained by sampling the physical transitions of neighboring states in the form of ($s$, $s'$, $t_{\rm Residence}(s)$).
\textbf{d}, Workflow for constructing the timescape dataset, $\mathcal{D} = \{(s, s', t_{\rm Residence}(s))\}$.
}
    \label{fig:scheme}
\end{figure*}

In the potential energy landscape of CrCoNi, a potential energy local minimum is denoted as a state $s$, which corresponds to a specific arrangement of atoms and vacancies in the face-centered cubic (\textit{fcc}) lattice sites. 
From a given state $s$, multiple atomic transitions can occur involving vacancy hopping to adjacent states $s'$. In this work, we define a mono-vacancy hop to a neighboring atomic site as an action, denoted by $a_{s\to s'}$. Each action corresponds to a transition rate, $\Gamma (a_{s\to s'})$, which can be evaluated by the TST~\cite{truhlar1996current}:
\begin{equation}
    \Gamma (a_{s\to s'})  = \nu_{\rm A} (s\to s') e^{-E_{\rm A}(s\to s')/k_{\rm B}T}
    \label{eq:TransitionRate}
\end{equation}
where $E_{\rm A}(s\to s')$ and $\nu_{\rm A} (s\to s')$ are the energy barrier and the attempt frequency, respectively. 
Then the average residence time of the system in state $s$ can be defined as:
\begin{equation}
    t_{\rm Residence}(s) = \frac{1}{\sum_{a\in \mathcal A_s} \Gamma (a)}.
    \label{eq:ResidenceTime}
\end{equation}
The set of all possible actions in the state $s$ constitutes the action space $\mathcal A_s$, while the collection of all possible states defines the state space $\mathcal S$. 

For a given temperature history $T(t)$, CrCoNi evolves into certain orderings of atomic configurations that, in the long-time constant-$T$ limit, minimize the free energy for that $T$. 
Thermodynamic equilibrium of SRO (SRO$^{\rm Eq}$) may be attained from this relaxation process if one waits for ``enough" transitions under constant $T$. 
We define the states corresponding to thermodynamic equilibrium as ``target states" ($\mathcal G$). 
The use of a set of states, rather than a single one, 
is necessary because in a simulation with $N^{\rm sim}$ atoms at finite temperature, equilibrium fluctuations scale as $1/\sqrt{N^{\rm sim}}$ on a per-atom basis~\cite{frenkel2023understanding}. 
Thus, the $\mathcal G$ ``volume" in phase space is finite and the boundary of $\mathcal G$, $\partial \mathcal G$, may be roughly estimated by a hyperspace distance cutoff to the centroid of $\mathcal G$.

In a physical process, a series of consecutive states forms a trajectory $\mathcal T$:
\begin{equation}
    \mathcal T \equiv (s_0, s_1, \cdots , s_k,\cdots , s_K)
\end{equation}
where $s_0$ is an initial atomic structure such as a completely random solid solution (RSS) state and $K$ is the time horizon of the trajectory to reach the hypersurface of $\mathcal G$ (Fig.~\ref{fig:scheme}a). 
It is desirable to track the kinetics of SRO along the evolution trajectory during the thermal annealing process. However, the SRO relaxation process may occur at macroscopic timescales at low temperatures, making direct kinetics simulations challenging. Thus, we develop the MFPT approach to address the long-timescale problem. 
The definition of the MFPT is an average time of trajectories ($t_{\mathcal T}$) that reach the $\mathcal G$, expressed as follows:
\begin{equation}
\begin{aligned}
    \text{MFPT}(s_0) &\equiv \sum_{\mathcal{T} \in \text{FP}(s_0)} P(\mathcal{T}) \, t_{\mathcal{T}}.
    \label{eq:mfpt}
\end{aligned}
\end{equation}
Here, $\text{FP}(s_0)$ is a set of first passage trajectories that begin from $s_0$ and first enter the $\mathcal G$ at its last state $s_K$, $t_{\mathcal{T}}$ is the average simulation time, and $P(\mathcal T)$ is the total probability of each trajectory. 
More details on the general expression of MFPT are provided in Sec. S(A).
It is evident that MFPT($s$) is equal to zero if $s\in \mathcal G$. Assuming ergodicity of the kinetic system, we have $\sum_{\mathcal T\in \text{FP}(s_0)} P(\mathcal T)=1$, as all trajectories enter $\mathcal G$ at least once if $K\to \infty$. 

The issue at hand is akin to ``how long does it take for a raindrop in the Rocky Mountains to end up in an ocean?".  To answer this question, one first has to determine {\em which} ocean: the Atlantic or the Pacific. It is perhaps unnecessary to track every rivulet or detail of every river bend to determine this region of final destination.
Once the region of final destination (\textit{e.g.}, SRO$^{\rm Eq}(T)$) has been identified for a given $T$, the next step is to estimate the {\em average} time of arrival. 
In principle, MFPT($s$) can be evaluated by sampling the $t_{\mathcal T}$ directly as shown in Fig.~\ref{fig:scheme}b. 
However, this approach becomes computationally prohibitive as the number of possible paths and $K$ increases.

In this paper, we take advantage of the Markovian network~\cite{LiKELQMDY11} assumption in developing the governing equation for MFPT. 
As described in Fig.~\ref{fig:scheme}c, we do not compute MFPT by sampling $\mathcal T$s because there are too many of them. 
Instead, we define a unique value of MFPT($s\rightarrow \mathcal G$) for any state $s$, and write down the recursive equation:
\begin{equation}
    \text{MFPT}(s\rightarrow \mathcal G) = t_{\rm Residence}(s) + 
    \sum_{a\in \mathcal A_s} P(a) \text{MFPT}(s'\rightarrow \mathcal G)
    \label{eq:MFPT}
\end{equation}
where $P(a)$ is the physical probability of taking a certain next-step action $s\rightarrow s'$ according to the TST in Eq.~\eqref{eq:TransitionRate}, with $\sum_{a\in \mathcal A_s} P(a)=1$, and $\text{MFPT}(s'\rightarrow \mathcal G)$ is the MFPT of the hopped-to state.  Note that Eq.~\eqref{eq:MFPT} is a self-consistent equation: given an initial guess of MFPT timescape, one can evaluate the right-hand side of the equation and obtain a new MFPT timescape. Such updates of timescape can be done iteratively until convergence. 
It is easy to see why Eq.~\eqref{eq:MFPT} may be more efficient for computing MFPT than naive trajectory sampling. As illustrated in Fig. S1, $1\leftrightarrow 2$ is the simplest example of the repeated crossing problem and can waste a lot of {\em computing resources} in naive kMC path sampling, but in the analytical equations for $\text{MFPT}(1)$, $\text{MFPT}(2)$, $\text{MFPT}(3)$ (we ignored the ``$\rightarrow\mathcal G$" notation when the final destination is clear), such repeated crossings effect is {\em entirely} captured by the recursive analytical equation and does {\em not} constitute extra computational cost, vis-a-vis the analytical treatment for any other node(s) as the small escape rate of $2\rightarrow 3 \rightarrow \mathcal G$ is the key for every node.  
Finally, we parameterize the timescape of $\text{MFPT}(s\rightarrow\mathcal G)$ by DNN, and solve for self-consistency in Eq.~\eqref{eq:MFPT} iteratively. 
The self-consistent timescape provides the answer to the ``how long does it take for a raindrop to end up in an ocean" question. 
The phrase ``timescape" can be appreciated by thinking about driving options in busy traffic hours, with a fixed final destination in mind. In the driving example, the self-consistency of the timescape comes from the cumulative properties of the driving time.


\subsection{General computational framework}\label{general_framework}  
Solving Eq.~\eqref{eq:MFPT} alleviates the need for direct sampling of $\lbrace{\mathcal{T}\rbrace}$ while preserving kinetic fidelity, but still requires sufficient coverage of states along $\mathcal{T}$.
As MFPT estimation involves traversing the state space reversely, from a state in $\mathcal{G}$ back to the initial state, explicit sampling of all possible states remains computationally intractable.  
To overcome this challenge, we adopt the TD learning that is formulated analogously to solving the Bellman Equation in RL.
Here, the MFPT timescape replaces the value function, and the recursive structure of Eq.~\eqref{eq:MFPT} enables iterative updates of the model by taking advantage of the expressive power of DNN in representing $\text{MFPT}_\theta(s)$, where $\theta$ stands for neural network weights. The global updates of $\theta$, performed through standard neural-network back-propagation, ``evolve" the timescape in an iterative manner. 
To construct a training dataset  $\mathcal{D} \equiv \{(s, s', t_{\rm Residence}(s))\}$, two subtasks are required: (i) identifying the thermodynamic destination $\mathcal{G}$, and
(ii) sampling physically plausible transitions along $\mathcal{T^{\rm TKS}}$ (Fig.~\ref{fig:scheme}d). The first guarantees that MFPT predictions are learned with respect to a well-defined destination, while the second ensures sufficient state-space coverage consistent with the underlying kinetics. 

To expedite the dataset construction, we employ the DRL-accelerated atomistic simulations (Fig. S2)., where the DRL agent serves as an accelerated MC sampler for free-energy minimization and as a kMC simulator for transition kinetics, corresponding to \textit{lower-energy states sampler} (LSS) and \textit{transition kinetics simulator} (TKS), respectively. 
Compared to standard MC annealers, LSS provides a computationally efficient approach for searching thermodynamic equilibrium states following a step change in an external condition, \textit{e.g.} relaxation, which considers the variation of free energy landscape at a longer time horizon. \cite{tang2024reinforcement}. Notably, as the discounting factor $\gamma$ increases, long-term optimization is prioritized, which contributes to reducing the number of simulation steps required to reach the thermodynamic destination $\mathcal{G}$ (Fig.~S3).

To determine the thermodynamic destination of SRO evolution (SRO$^{\rm Eq}$), we combine DRL-LSS, which evolves the system with physically realizable vacancy hops, and Metropolis Monte Carlo (MMC) sampling~\cite{metropolis1953equation}, which allows unphysical atomic swaps. The trajectories ($\mathcal T^{\rm DRL\text{-}LSS}$ and $\mathcal T^{\rm MMC}$) of both simulations are driven by free-energy minimization and thus converge to $\mathcal G$. 
Although MMC introduces nonphysical and often sparse evolution trajectories, it remains valuable for accurately estimating the centroid and fluctuation width of $\mathcal{G}$.  
Ultimately, we aim to extend DRL-LSS to accurately estimate both the equilibrium center and its fluctuations using only physically valid vacancy hops, eliminating the need for unphysical sampling.

Having obtained trajectories to $\mathcal{G}$ from DRL-LSS and MMC, we employ the TKS to generate kinetics-informed trajectories,  $\mathcal{T}^{\rm TKS}$, starting from randomly selected intermediate states along  $\mathcal T^{\rm DRL\text{-}LSS}$ and $\mathcal T^{\rm MMC}$. Due to stochastic nature of diffusion, these TKS-generated trajectories encourage systematical exploration of transition networks from vacancy hopping events governed by kinetic diffusion barriers, offering sufficient data for training.

 
In this work, we develop three key models: a reaction model, a deep Q-network (DQN), and a timescape estimator $\text{MFPT}_\theta(s)$, which require an atomic structure $s$ or a state transition $(s,s’)$ as inputs (Fig. S4).
The reaction model is used to generate $\mathcal T^{\rm TKS}$ by efficiently predicting microscopic potential energy barriers $E_{\rm A}(a_{s\to s'})$, and attempt frequencies  $\nu_{\rm A}(a_{s\to s'})$, of vacancy hopping events without performing computationally intensive Nudged-Elastic Band (NEB) calculations. 
Next, we train the DQN for the DRL-LSS to obtain an optimal transition policy that samples an action for any given state  $s$ to approach $\mathcal G$ more efficiently.
Finally, we solve the self-consistent timescape equation Eq.~\eqref{eq:MFPT} by training the $\text{MFPT}_\theta(s)$.
 In total, with Sec.~\ref{secReactionEncoding} providing microscopic information about hopping barriers and DRL accelerated atomistic simulation, Sec.~\ref{secSROeq} providing the thermodynamic destination and Sec.~\ref{secMFPT} providing the timescape estimation, we can build a bridge to thermodynamic equilibrium with tracking of the physical time.

\subsection{DRL atomistic simulation of vacancy hopping}
\label{secReactionEncoding}

\begin{figure}[!htb]
    \centering
    \includegraphics{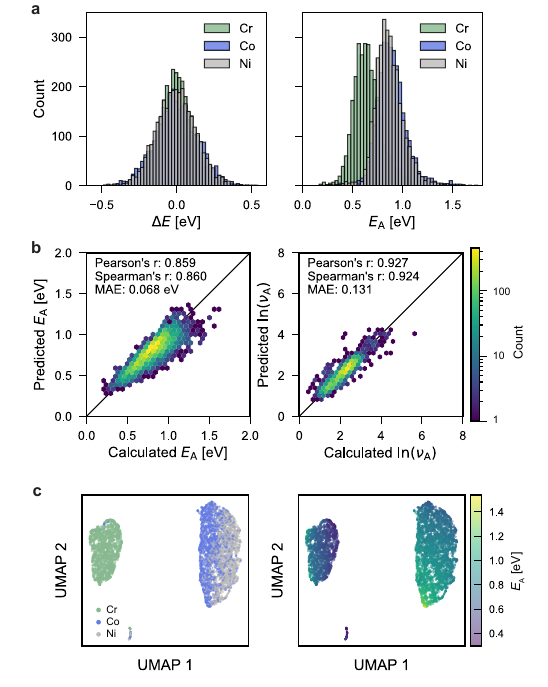}
    \caption{Encoding atomic transitions and reaction model performance.
    \textbf{a}, Histogram of the energy changes ($\Delta E$) and energy barriers ($E_{\rm A}$) for vacancy diffusion events involving each type of atoms.  
    \textbf{b}, Prediction accuracy on $E_{\rm A}$, and attempt frequency $\ln(\nu_{\rm A})$ of the GNN-based reaction model for in-distribution test set. 
    \textbf{c}, Distribution of the projected reaction encoding vector $\mathcal R$ in the training data. The left panel is colored by the hopping element type and the right one is colored by the corresponding $E_{\rm A}$ values.
    }
    \label{fig:new_model}
\end{figure}
We leverage geometric deep learning with GNNs to encode atomic structures and vacancy hopping transitions.  
To calculate the ``ground truth" potential energy of the system as well as $E_{\rm A}$ and $\nu_{\rm A}$ values for vacancy hopping events, we utilize a universal ML-IP, \texttt{MACE-MP-0}~\cite{batatia2023foundation}. 
Details on dataset creation, model architecture, and training process are provided in Sec.~\ref{reaction_model_method}.
Figures~\ref{fig:new_model}a and S5 show the distribution of the potential energy change $\Delta E$, the energy barrier $E_A$, and the attempt frequency $\ln(\nu_{\rm A})$. The energy change ($\Delta E$) exhibits a unimodal distribution centered around zero for all elements. The activation energies computed by the universal ML-IP range within three standard deviations of the mean, from 0.268 to 1.333~eV, in good agreement with the previous reports~\cite{du2022chemical,zhao2018effect}. Interestingly, activation energies for vacancy diffusion to Cr sites are notably lower compared to those for Co and Ni sites, suggesting a higher mobility of Cr compared to the other two elements, which provides a practical way to tune SRO formation kinetics through varying Cr concentration. 
On the other hand, the vacancy concentration $X_{\rm vac}$ may have little influence on thermodynamic quantities such as SRO$^{\rm Eq}$, because $X_{\rm vac}^{\rm Eq}$ is dilute (typically ranging from $10^{-7}$ to $10^{-9}$). In other words, unlike the major alloying elements like $X_{\rm Cr}$, which are both thermodynamically and kinetically important, it is generally well accepted and understood that while $X_{\rm vac}$ may be key for kinetics following the Kirkendall-Smigelskas experiment in 1947~\cite{nakajima1997discovery}, it can often be ignored for thermodynamic equilibrium properties due to dilute concentrations.

The GNN-based reaction model successfully learns the thermodynamics and kinetics of vacancy hopping events and demonstrates improved predictive accuracy compared to the previous model~\cite{tang2024reinforcement} (Fig.~\ref{fig:new_model}b and S6-9). More details and quantitative comparisons are provided in Sec. S(B) of the Supplementary Information.
Furthermore, the learned reaction encoding vector, $\boldsymbol{\mathcal R}$, effectively captures the underlying features of vacancy hopping in CrCoNi. 
Figure~\ref{fig:new_model}c presents 2D projections of $\boldsymbol{\mathcal R}$ using the Uniform Manifold Approximation and Projection (UMAP) algorithm~\cite{mcinnes2018umap}, which reveals three distinct clusters, with Co and Ni generally residing in the same cluster, while Cr forms a separate cluster. Interestingly, the reaction encoding vector $\boldsymbol{\mathcal R}$ is able to further distinguish Co and Ni within the same cluster. The color-coding of the activation energies also aligns with their distribution, highlighting that vacancy hops to Cr sites exhibit lower activation energies compared to those involving Co or Ni.

The GNN-based approach, which automatically learns representations for both structural and reaction inputs, offers several advantages. 
Firstly, the model exhibits good prediction generalizability on large atomistic systems because of the input parametrization with atomic feature difference. Due to the localized vacancy hoppings during SRO formation, atoms distant from the diffusion event exhibit minimal changes in their local environments, resulting in near-zero atomic features difference (before and after vacancy hopping) for these atoms. 
This locality enables the reaction model to scale up to larger cells efficiently as long as the focus remains on local events, since near-zero atomic feature difference of distant atoms barely contribute to reaction model predictions. 
Additionally, trained encoding vectors of structures and reactions in the reaction model serve as a warm start, providing a good pre-trained representation (input features) for the DRL framework and the timescape estimator, thereby enhancing training efficiency and convergence.
Finally, the framework is not strictly limited to GNNs but can be applied with any model capable of representing transitions.
Although the proposed method is particularly effective for local atomic transitions, alternative approaches, such as transformers, may be required to encode long-range transitions and interactions. 

With these advantages, we integrate the reaction model into the DRL framework to simulate SRO formation in CrCoNi. On the one hand, the reaction model can directly predict both the $E_{\rm A}$ and $\ln(\nu_{\rm A})$ values that are used in TKS to generate kinetics-informed trajectories $\mathcal T^{\rm TKS}$.
On the other hand, the DQN is initialized with the reaction model and further trained to construct the $\mathcal T^{\rm DRL-LSS}$. As shown in Fig.~S10, DRL-LSS demonstrates substantially higher computational efficiency for energy minimization and SRO evolution than both kMC and MMC when constrained to the same action space of physical vacancy hops. 
More details on reward design, DQN training, and performance evaluation are provided in Sec.~\ref{dqn_method} and Supplementary Sec.~S(C).

\subsection{Target states of SRO formation}
\label{secSROeq}
\begin{figure*}[!htb]
    \centering
    \includegraphics{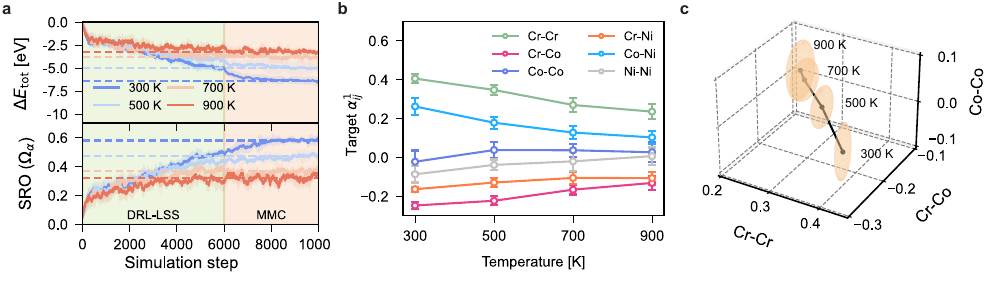}
    \caption{Target states ($\mathcal G$) of SRO formation. 
\textbf{a}, SRO formation in CrCoNi with 0.39~at\% vacancy concentration at various temperatures. The first 6,000 simulation steps are performed using the DRL-LSS with a simulated annealing schedule, defined as $T = 1200 - \frac{1200 - T_{\rm anneal}}{4500} \times {\rm step}$ over 4500 simulation steps, followed by equilibration at a constant annealing temperature $T_{\rm anneal}$ for an additional 1500 steps. The final configuration from the DRL-LSS is then further equilibrated using MMC sampling at $T_{\rm anneal}$ to determine the centroid and fluctuation of the target states. Dashed lines indicate the ${\rm SRO^{\rm Eq}}(T)$, averaging the last 1,000 frames.
\textbf{b}, Temperature-dependent evolution of WC parameters ($\alpha_{ij}$).  
\textbf{c}, Temperature-dependent ellipsoidal target states. The lines connect the centers of the $\mathcal{G}(T)$ across different temperatures, while the ellipsoidal volumes represent equilibrium fluctuations, characterized by $1\sigma^{\rm SRO}_{q}$.
}
    \label{fig:target_states}
\end{figure*}

As described in Sec.~\ref{general_framework}, we define the set of ``target states'', $\mathcal G$, for SRO formation as atomic configurations that lie within the thermodynamic equilibrium fluctuations of the vacancy-mediated diffusive relaxation.
To characterize the centroid and the extent of these fluctuation, we adopt the Warren–Cowley (WC) parameter~\cite{cowley1965short}, which quantifies the likelihood of finding element pairs $i$ and $j$ as $n^{\text{th}}$-nearest neighbors, defined as $\alpha_{ij}^{n} \equiv 1 - \frac{p_{ij}^{n}}{c_i c_j}.$, where $p_{ij}^{n}$ is the probability of observing an $i$–$j$ pair in the $n^{\text{th}}$-nearest neighbor shell, and $c_i$ and $c_j$ are the concentrations of elements $i$ and $j$, respectively. 
If elements $i$ and $j$ are randomly distributed, $\alpha_{ij}^{n} \approx 0$, and negative (positive) values indicate a tendency for attraction (repulsion) between the elements. 
In this work, the overall degree of SRO is quantified using the Euclidean norm of the first nearest neighbor WC parameters ($\Omega_{\alpha} \equiv \sqrt{\sum_{i \leq j} |\alpha^1_{ij}|^2}$). 
The centroid of $\mathcal{G}$ is given by the equilibrium average of the first nearest neighbor WC parameters $\overline{\alpha_{ij}^{1}}$ while the extent of equilibrium fluctuations is captured by $\sigma^{\mathrm{SRO}}_{q} \equiv \sqrt{\text{Eigvals}(\boldsymbol{\Sigma})_q}$, with $\boldsymbol{\Sigma}$ being the covariance matrix of the WC parameters at equilibrium and $\text{Eigvals}(\cdot)_q$ means the $q^{\text {th}}$ eigenvalue of the matrix. 
Isotropic SRO fluctuations yield similar $\sigma^{\mathrm{SRO}}_{q}$ values across different element pairs, while anisotropic fluctuations manifest pair-dependent variations in $\sigma^{\mathrm{SRO}}_{q}$. In this work, the $\mathcal{G}$ is defined as the set of configurations within $2\sigma^{\mathrm{SRO}}_{q}$ fluctuations to ensure a sufficient fraction of data points satisfying $s\in \mathcal G$ included in the training set.

Figure~\ref{fig:target_states}a illustrates the change of potential energy from an initial state and SRO with the number of simulation steps.
The DRL-LSS simulation rapidly relaxes the system towards thermodynamic equilibrium, as indicated by the potential drop, through physically valid vacancy hopping events, while subsequent MMC sampling rigorously equilibrates the system and identifies the target state $\mathcal{G}$. During the process, the degree of chemical order incrementally increases, which is favored by reduced enthalpy from complicated solute-solute interactions.
As temperature decreases, the system exhibits a higher $\Omega_{\alpha}$ value accompanied by reduced thermal fluctuations, reflecting enhanced thermodynamic ordering resulting from decreasing influence of configurational entropy at low temperatures (Fig.~S11).
Figure~\ref{fig:target_states}b and Table~S1 further present the temperature dependence of the first-nearest-neighbor Warren–Cowley parameters $\alpha^{1}_{ij}(T)$ of the element pairs at different temperatures.
Across a wide range of temperatures between 300 K and 900 K, Cr tends to repel itself while attracting Co and Ni, whereas Co and Ni mutually repel, as evidenced by the positive parameters of Cr-Cr, Co-Ni pairs and negative values between Cr-Co, Cr-Ni pairs , which are consistent with the previous findings~\cite{cao2024capturing}.

Figure~\ref{fig:target_states}c shows the temperature dependence of the target states,  $\mathcal G(T)$, in the SRO parameter space. 
As the temperature decreases, the centroid of $\mathcal G$ drifts towards regions with larger WC values, indicating a stronger degree of chemical ordering, where the drift path is depicted by the solid line connecting the ellipsoid centroids. 
Additionally, the orientation and shape of $\mathcal G$ also evolve with the temperature change resulting from the corresponding rotation and dilation of the target states. 
For example, as the temperature decreases from 500 K to 300 K, one can observe an elongation along Co-Co axis and contraction Cr-Cr, Cr-Co axes (dilation), as well as a slightly tilt towards Co-Co axis (rotation). 
Interestingly, the fluctuation anisotropy becomes increasingly pronounced at lower temperatures, with the dominant elongation emerging along the Co–Co axis. This behavior reflects the fact that Co–Co correlations strengthen more rapidly upon cooling than the Cr–Cr or Cr–Co modes, leading to a softer curvature of the free-energy landscape in the Co–Co direction and consequently larger thermal fluctuations along that axis. At sufficiently high temperatures, the fluctuation volume approaches an isotropic sphere centered at the origin (assuming the material remains solid), consistent with random-walk statistics. Conversely, the fluctuation amplitude vanishes as the temperature approaches absolute zero.


\subsection{Learning the timescape of SRO formation}
\label{secMFPT}

\begin{figure}[!htb]
    \centering
    \includegraphics{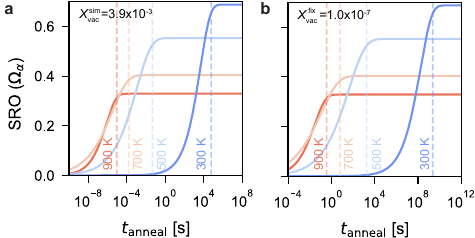}
   \caption{
The SRO evolution times based on the $\rm {MFPT}$ prediction at various annealing temperatures.  
\textbf{a}, With a vacancy concentration from the simulation cell $X_{\rm vac}^{\rm sim}$.  
\textbf{b}, With a fixed dilute vacancy concentration $X_{\rm vac}^{\rm fix}$. Dashed lines represent the predicted annealing time $t_{\rm anneal}$ when the overall degree of SRO $\Omega_{\alpha}$, has completed 95\% of its approach to the plateau value.
}
    \label{fig:time_deploy}
\end{figure}

As introduced in Sec.\ref{conceptual_framework} and Sec.\ref{general_framework}, our goal is to estimate the MFPT from any state $s$ at a given temperature $T$ using a DNN, answering the question: ``How long does it take for a raindrop to end up in an ocean?". 

To obtain the MFPT, we design the timescape estimator, $\text{MFPT}_\theta(s, X_{\rm vac}, T)$, to take as input the vacancy concentration $X_{\rm vac}$, temperature $T$, and atomic configuration $s$, enabling prediction across different processing conditions. 
To improve accuracy and generalization, we leverage the learned atomic representations from the pre-trained reaction model and apply a multitask learning approach that simultaneously estimates MFPT and classifies whether $s \in \mathcal{G}$.
Additional details on the training procedure and evaluation results are provided in Sec.~S(D) and Supplementary Sec.~S(E).
To map the MFPT to the kinetics of SRO formation during thermal annealing, we define the annealing time for a given state $s$ as
\[
t_{\mathrm{anneal}}(s) \equiv \text{MFPT}_\theta(s_0) - \text{MFPT}_\theta(s),
\]
where $s_0$ denotes the initial RSS configuration and $\text{MFPT}_\theta(s_0)$ denotes the corresponding MFPT to reach $\mathcal{G}(T)$. 
We then fit the SRO  as a function of annealing temperature and $t_{\rm anneal}$ using a parametric model.
Further details on this mapping are provided in Supplementary Sec.~S(F).

Fig.~\ref{fig:time_deploy} illustrates the variation of the $\Omega_{\alpha}$ in CrCoNi as a function of the annealing time at different temperatures and vacancy concentrations. In Fig.~\ref{fig:time_deploy}a, with a vacancy concentration, $X_{\rm vac}^{\rm sim}$, as $3.9\times 10^{-3}$, the equilibrium SRO exhibits a gradual increase at decreasing temperatures as in Fig. 3a. Importantly, the corresponding annealing time spans ten orders of magnitude, increasing significantly from $10^{-5}\rm s$ at 900 K to $10^{5}\rm s$ at 300 K. As shown in Fig.~S9, the MFPT predictions closely match the TKS kinetics in the high-temperature regime (\textit{e.g.}, 900~K). At lower temperatures, however, the SRO evolution in TKS no longer reaches SRO$^{\rm Eq}(T)$ within feasible simulation time, whereas the MFPT prediction successfully approaches SRO$^{\rm Eq}(T)$ and predicts substantially longer longer annealing times. These results signify that the MFPT prediction not only reproduces the accessible high-temperature kinetics, but also enables reliable extrapolation into low-temperature regimes where conventional kMC simulations become computationally prohibitive.

To further assess the physical timescale relevant to materials processing where the vacancy concentration can be dilute, we extrapolate our predictions of $t_{\rm anneal}$ under the assumption that the relaxation timescale is inversely proportional to the vacancy concentration. Fig.~\ref{fig:time_deploy}b presents the dependence of $\Omega_{\alpha}$ on the annealing time at the four temperatures where the vacancy concentration, $X_{\rm vac}^{\rm fix}$, is set as $1\times 10^{-7}$ corresponding to the dilute concentration. Compared to $X_{\rm vac}^{\rm sim}$, the annealing time increases by around four orders of magnitude at all temperatures, manifesting the critical role of vacancy concentration in governing SRO formation during thermal annealing.


\begin{figure}[!htb]
    \centering
    \includegraphics{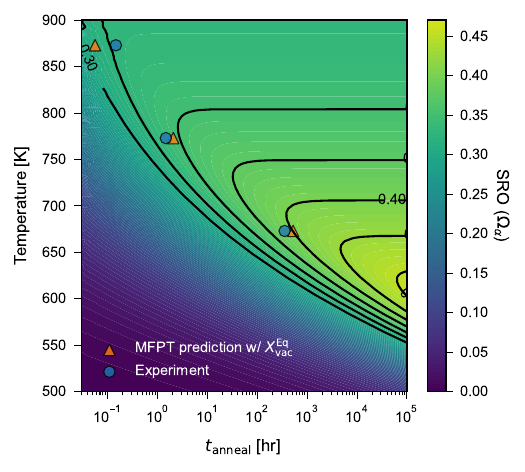}
    \caption{Structure–process relationship of SRO formation in CrCoNi. Time–temperature–SRO map under isothermal annealing at $T_{\rm anneal}$ assuming a fixed vacancy concentration of $X_{\rm vac}^{\rm Eq}(T_{\rm anneal} )$. 
    Scatter points indicate the predicted annealing time $t_{\rm anneal}$ at which the SRO reaches 95\% of its total evolution toward the plateau, as estimated by our model. For comparison, experimental measurements~\cite{li2023evolution} of electrical resistivity, which serve as a proxy for SRO evolution, are also included.
}
    \label{fig:time_map}
\end{figure}

It is worth noting that vacancy concentrations are fixed across the temperatures to estimate the annealing time in Fig.~\ref{fig:time_deploy}. In reality, however, the equilibrium vacancy concentration $X_{\rm vac}^{\rm Eq}(T)$ is temperature dependent. Previously, the equilibrium vacancy concentration $X_{\rm vac}^{\rm Eq}(T)$ was estimated using the average vacancy formation energy combined with configurational mixing entropy~\cite{du2022chemical}.
However, in multicomponent alloys, the vacancy formation energy strongly depends on the local atomic environment and therefore cannot be adequately represented by a single scalar value.
An alternative approach considers the distribution of local vacancy formation energies to obtain an effective vacancy formation energy~\cite{li2024vacancy}.
In recent work, we introduced a grand canonical Monte Carlo (GCMC)-based gedankenexperiment that explicitly samples the spectral distribution of site-specific vacancy formation free energies, enabling direct computation of the equilibrium vacancy concentration as a function of temperature. In this study, we adopt those GCMC-derived $X_{\rm vac}^{\rm Eq}(T)$ values for our analysis (Fig. S15).

To capture the accessible processing space for SRO emergence, we construct a time–temperature–SRO map as shown in Fig.~\ref{fig:time_map}.
These results highlight a key trade-off: higher annealing temperatures enable rapid equilibration, but limit the attainable degree of ordering, while lower temperatures promote stronger ordering at the expense of prohibitively long annealing times. 
We further validate our predictions using experimental measurements of electrical resistivity, which serve as a proxy for the evolution of SRO~\cite{li2023evolution, pfeiler1988investigation, pfeiler1988investigation} with the linear relationship. Due to the linear relationship, the increase in electrical resistivity in the experiment reaches 95\% of its saturation value at the same time as the SRO reaches 95\% of its saturation value. Therefore, the experimental observations of the annealing time at 95\% of electrical resistivity can be directly compared to the MFPT prediction of the annealing time at 95\% of the SRO formation. More detailed discussion on the relationship between SRO and electrical resistivity is provided in Supplementary Sec.~S(F).
Future work may explore a more direct property–processing relationship between electrical resistivity and annealing time.
The scatter plots in Fig.~\ref{fig:time_map} show good agreement between the MFPT predictions and experimental observations, particularly in the low-temperature regime.
For example, at 673~K, the predicted annealing time of 540~hours is reasonably consistent with the experimentally measured value of 352~hours.
It also reveals that the required annealing times at temperatures below 500~K are predicted to exceed practical timescales relevant to industrial or experimental settings, highlighting the intrinsic kinetic limitations of equilibrium SRO formation at low temperatures.


\section{Discussion}
The proposed strategy of DRL-accelerated atomistic simulation and timescape learning offers a generalizable framework for estimating the long-timescale kinetics and elucidating processing-structure-property relation.
By mapping atomic-scale structure (\textit{e.g.}, SRO) relaxation dynamics to a target property such as mechanical strength, magnetism, and thermal stability, this approach enables predictive control over material performance. 
Importantly, our framework circumvents the intrinsic timescale limitations of conventional atomistic simulations, extending its applicability to broader classes of slow relaxation phenomena, such as grain boundary evolution~\cite{MishinAL10} and other long-timescale structural transformations in materials.

Nevertheless, important opportunities remain for further refinement and broader validation. The prediction error primarily arises from two competing factors. The first is the equilibrium fluctuation of $2\sigma^{\mathrm{SRO}}_{q}(T)$, which leads to an underestimation of the annealing time. The second is that the vacancy concentration is fixed to its equilibrium value at the annealing temperature, \textit{i.e.}, $X_{\rm vac}^{\rm Eq}(T_{\rm anneal})$.
This simplification neglects the transient evolution of the vacancy population during annealing and assumes instantaneous equilibration.
As a consequence, the predicted annealing time $t_{\rm anneal}$ may be overestimated, particularly in regimes where vacancy diffusion is slow and $X_{\rm vac}$ deviates from its equilibrium value over experimentally relevant timescales.
At high temperatures, the equilibrium vacancy concentration is close to that at elevated temperatures, making fluctuation-related effects dominant. In contrast, as the annealing temperature decreases, the equilibrium vacancy concentration decreases exponentially, and the overestimation effect becomes dominant.

A central area for improvement lies in the definition of the thermodynamic destination. 
While the 1$^{\rm st}$-order SRO quantification metric employed in this study effectively captured equilibration behavior, it may exhibit degeneracies that obscure distinctions between distinct ordering motifs. 
Incorporating higher-order WC parameters or alternative metrics could improve sensitivity and robustness in detecting convergence. 
Recent computational advances~\cite{sheriff2024quantifying} have demonstrated the utility of equivariant neural networks in accurately quantifying atomic ordering in multimetallic systems.
However, significant experimental challenges persist~\cite{he2024quantifying}, particularly in resolving lower-order nearest-neighbor correlations (\textit{e.g.}, below 7NN), which often rely on indirect proxies such as changes in physical properties~\cite{li2023evolution}. 
Future efforts can focus on establishing transferable descriptors that are consistent across both simulations and experiments, thereby enhancing the precision and reliability of SRO characterization.

Our findings highlight the critical role of vacancy concentration in determining the annealing timescales. 
In this study, we assumed that the vacancy concentration is equilibrated at the annealing temperature, thereby neglecting the pre-annealing homogenization step at elevated temperatures and the associated transient evolution of vacancies. 
The transient evolution of the vacancy concentration depends on detailed experimental conditions, including thermal histories and microstructural features such as grain size and polycrystallinity, which can consequently influence the kinetics of SRO formation. Incorporating this case-specific information from the target experiment can further improve the prediction accuracy of the SRO evolution.

\section{Methods}\label{methods}
\subsection{Reaction model}
\label{reaction_model_method}
GNNs are widely used to represent atomic structures, where atoms serve as nodes and connections between neighboring atoms as edges. A GNN trained with available property labels, such as total energy or atomic forces, can automatically learn the local atomic environments around a center atom. 
In this work, we leverage the deep geometric learning capabilities of GNNs to encode vacancy hopping transitions by computing the difference between the learned representations of two states, $s$ and $s'$, following a strategy similar to previous studies~\cite{grambow2020deep, coley2019graph}.
We employed the polarizable atom interaction neural network (PaiNN) model~\cite{schutt2021equivariant}, which ensures rotational equivariance for vector features within model layers, as our message-passing architecture.
The GNN model was initially trained on a simpler task: predicting the total energies $E^s$ of states $s$. 
In this task, the GNN also outputs a representation vector on each node denoted as $\boldsymbol{z}_n$ ($n$ is the index of the node).
Then, $\boldsymbol{z}_n$ was used to predict our target learning task: predicting quantities associated with atomic transitions. The differences between the atomic representations $\boldsymbol{z}_n$ of $s$ and $s'$ are then computed to generate the atomic difference features. These features, along with the predicted energy difference (after relaxation), $\Delta E_{\rm pred}$, where $\Delta E_{\rm pred}\equiv E^{s'}-E^s$, are aggregated into a reaction encoding vector, $\boldsymbol{\mathcal R}$, defined as $\boldsymbol{\mathcal R}\equiv\Delta E_{\rm pred}\oplus \text{SumPool}\left(\boldsymbol{z}_n^{s’} - \boldsymbol{z}_n^{s}\right)$.
Here, $\oplus$ stands for concatenation and SumPool is the sum operation for all atoms $n$ for each component of the atomic feature vector. 
The reaction encoding vector was then used to predict key reaction outputs, including activation energy ($E_{\rm A}$), and attempt frequency ($\nu_{\rm A}$) for a given vacancy hop. 

To pre-train the reaction model, we constructed model systems of CrCoNi containing 108, 256, and 500 atoms with 1 to 4 vacancies. The dataset included configuration pairs with associated labels of the total energy of two states, $s$ and $s’$, denoted as $E^s$ and $E^{s’}$, as well as $E_{\rm A}$ and $\nu_{\rm A}$. 
The $E_{\rm A}$ was calculated as the energy difference between the transition state, ${\rm TS}_{s\rightarrow s'}$ and the initial state, $s$. 
The $\nu_{\rm A}$ was determined using harmonic transition-state theory (HTST)~\cite{tang2024reinforcement}. Transition state geometries were obtained using the NEB method~\cite{jonsson1998nudged, henkelman2000improved}, as implemented in the Atomistic Simulation Environment (ASE)~\cite{larsen2017atomic}. 
The full dataset included 8,911 training samples, an in-distribution test set of 3,811 samples, and an out-of-distribution test set of 628 samples generated from 864-atom systems to evaluate model generalization.
The training loss function is defined below:
\begin{equation}
\notag
    \mathcal L=\mathcal L_{\Delta E}+\mathcal L_{E_{\rm A}}+\mathcal L_{\ln\nu_{\rm A}}+0.1(\mathcal L_{E_s}+\mathcal L_{E_{s'}}),
\end{equation}
where $\mathcal{L}_{\Delta E}$ is the loss for the predicted energy difference between states, $\Delta E \equiv E_{s’} - E_s$, $\mathcal{L}_{E_{\rm A}}$ is the loss for the activation energy, $\mathcal{L}_{\ln\nu_{\rm A}}$ is the loss for the natural logarithm of the attempt frequency, and $\mathcal{L}_{E_s}$ and $\mathcal{L}_{E_{s’}}$ are the losses for the total energy of the initial and final states, respectively, weighted by 0.1 to maintain focus on the energy differences and reaction parameters. All components of the loss were computed using mean square error (MSE). 
Model training was performed using the Adam optimizer with a learning rate of $3\times 10^{-4}$, a batch size of 8, and a total of 150 epochs.

\subsection{Deep Q-network}
\label{dqn_method}
The DQN is trained by optimizing the state-action value function $q(s,a)$, and is used to formulate a policy $\pi(a|s)$ based on the Boltzmann distribution of $q(s,a)$ to account for the temperature effect and stochasticity. Since actions are taken based on probabilities, the system's evolution is non-deterministic. The $q(s,a)$ is trained by maximizing the total reward ($R$), which is the accumulated reward ($r$) with a discount factor, $\gamma$, as defined in Eq.\eqref{eq:total_reward} and \eqref{eq:reward}:

\begin{equation}
R \equiv \sum_{k=0}^{K-1}\gamma^k r_k.
\label{eq:total_reward}
\end{equation}
\begin{equation}
\begin{aligned}
r_k \equiv \alpha r^{\rm TKS}_k + \beta r^{\rm LSS}_k\\
=  -\alpha\left(\tilde{F}^{\rm saddle}_k-F_k\right)-\beta\left(F_{k+1}-F_k\right)\\
\approx
-\left(\alpha E_{\rm A} + \beta \Delta E\right) + \alpha k_{\rm B}T\ln \nu_{\rm A},
\end{aligned}
\label{eq:reward}
\end{equation}
where $E_{\rm A}$ is the activation energy, $\Delta E$ is the reaction energy, and $\nu_{\rm A}$ is the attempt frequency, with $T$ being the temperature and $k_{\rm B}$ the Boltzmann constant.
The reward function, $r_k$, is a sum of the kinetic ($r^{\rm TKS}_k$) and thermodynamic ($r^{\rm LSS}_k$) contributions of free energy change of an action, $a_{k\to k+1}$ , weighted by coefficients $\alpha$ and $\beta$.
Here, the kinetic reward term $r^{\rm TKS}_k$ accounts for the reaction rate, quantified by the difference between the free energy at the saddle point ($\tilde{F}^{\rm saddle}_k$) and the free energy of the initial state ($F_k$). The thermodynamic reward term $r^{\rm LSS}_k$ captures the free energy difference between two states, $F_{k+1}$ and $F_k$. This formulation accounts for both thermodynamic and kinetic driving forces, as well as future rewards, thereby accelerating the simulation process~\cite{tang2024reinforcement}.
In this work, we approximate the free energy difference $F_{k+1}-F_k$ of the stable states as the total energy difference, $\Delta E$. Also, quantum effects such as zero-point energies are not considered as they are negligible for heavy atoms systems at the temperature range of our study.

The learnable parameters of the DQN are updated according to the Bellman optimality equation:
\begin{equation}
\theta \leftarrow \theta - \lambda \nabla_\theta \sum_k \left[ r_k + \gamma \max_{a’} q_{\rm target}(s_{k+1}, a’) - q_\theta(s_k, a) \right]^2.
\end{equation}
Here, $q_{\rm target}$ is a target network that updates $\theta$ less frequently, and $\lambda$ is the learning rate. Note that the output of the DQN does not directly correspond to a physical quantity; instead, it represents the expected cumulative reward associated with a particular action. As a result, an action $a$ is selected from the learned policy $\pi_\theta(a|s)$, which follows the Boltzmann distribution $\pi_\theta(a|s)=\frac{\exp(q_\theta(s,a)/k_{\rm B}T)}{\sum_{a'\in\mathcal A_s}\exp(q_\theta(s,a')/k_{\rm B}T)}$, to introduce stochasticity and applied to the current state $s$, resulting in a transition to the next state $s'$. 
For certain conditions of interest, we can use a simplified version. If we focus solely on the instantaneous kinetic reward, setting $\beta = 0$ and $\gamma = 0$, the DRL framework is used as the TKS. Here, directly using the reaction model is sufficient. Alternatively, setting $\alpha = 0$ and $\gamma > 0$ emphasizes long-term free energy minimization but ignores the physical reaction probabilities, thereby breaking the kinetic fidelity of the atomic transitions.

For DQN training, the representation layers are initialized using the pre-trained weights from the reaction model, providing a warm start.
The DQN feature vector, $\boldsymbol{\rm x}_{\rm DQN}$, is formulated as $\boldsymbol{\rm x}_{\rm DQN} \equiv q_0 \oplus q_1 \cdot k_{\rm B}T \oplus \boldsymbol{\mathcal R}$, where $q_0 \equiv -\alpha E_{\rm A} - \beta \Delta E$ and $q_1 \equiv \alpha \cdot \ln\nu_{\rm A}$.
This composite feature vector $\boldsymbol{\rm x}_{\rm DQN}$ is used as input to the neural network to predict $q_{\theta}$. 
During the DQN training for DRL-LSS, we adopted $\alpha=0$, $\beta=1$, and $\gamma=0.8$. 
The time horizon was set to 30, and 50 episodes were performed starting from independent initial atomic configurations.
\subsection{Atomistic simulation}
In this work, we employed three types of atomistic simulations: MMC, TKS, and DRL-LSS. 
For MMC, we investigated two different action spaces: one involving vacancy hopping to first-nearest neighbor sites, 
and the other including unphysical element-swap actions. 
Equiatomic $\mathrm{Cr_{85}Co_{85}Ni_{85}}$ structures containing a single vacancy were randomly generated as the initial RSS configurations, ensuring that the initial SRO was below 0.03. 
To minimize statistical sampling errors and bias from initial structure selection, 
five independent simulations were performed for each temperature. 
For the TKS simulations, we used a DQN model with $\alpha = 1$, $\beta = 0$, and $\gamma = 0$, 
which predicts the attempt frequency and activation energy to accelerate the kMC simulation.


\subsection{Timescape  estimator}\label{time_estimator}
The input feature vectors in the timescape estimator,denoted as $\boldsymbol{\rm x}_{\rm time}$, incorporates atomic structure encoding vectors along with vacancy concentration and temperature information. It is formulated as $\boldsymbol{\rm x}_{\rm time}=X_{\rm vac}\oplus k_{\rm B}T\oplus {\rm MeanPool}(\boldsymbol{z}_{n}^s)$. MeanPool is the mean operation for all atoms $n$ for each component of the atomic feature vector. Finally, $\rm MFPT_\theta(s)$ for a given vacancy concentration $X_{\rm vac}$ and temperature $T$, is evaluated as:
\begin{equation}
{\rm MFPT}_\theta(s, X_{\rm vac}, T) = -\tau_{X_{\rm vac}, T} \cdot \ln\left(1 - \frac{t_\theta(s)}{\tau_{X_{\rm vac}, T}}\right),
\end{equation}

The model was trained using the Adam optimizer with a learning rate of $1 \times 10^{-5}$.
The training was carried out with a batch size of 64 over 300 epochs.
Details on the selection of hyperparameters for $Q_{\rm A}$, $\tau_0$, and the loss weight coefficients ($\omega$ and $\omega_{\rm Cls}$) are provided in Supplementary Sec.~S(E).
To construct the timescape dataset, we randomly selected intermediate states from $\mathcal{T}_{\rm LSS}$, which served as starting points for subsequent TKS trajectories consisting of 30 or 50 steps across the temperatures.
As a result, we generated 65,500 data points. The total dataset was split into training (80\%), validation (2\%), and test (18\%) sets for training and evaluating the timescape estimator. Note that the validation set is for monitoring the training progress.


\section{Data Availability}
The trained models will be available in  \href{https://figshare.com/}{Figshare}. 
Our training and testing datasets are available upon reasonable request to corresponding authors.

\section{Code Availability}
The codebase to train models and run DRL-accelerated atomistic simulations is available at \href{https://github.com/learningmatter-mit/RLVacDiffSim}{https://github.com/learningmatter-mit/RLVacDiffSim}. The codebase for graph neural network based surrogate models is available at \href{https://github.com/learningmatter-mit/ReactionGraphNeuralNetwork}{https://github.com/learningmatter-mit/ReactionGraphNeuralNetwork}.

\section{Acknowledgements}
J.L. and H.T. acknowledge support by NSF DMR-1923976. This material is based upon work supported by the Under Secretary of Defense for Research and Engineering under Air Force Contract No. FA8702-15-D-0001. Any opinions, findings, conclusions or recommendations expressed in this material are those of the author(s) and do not necessarily reflect the views of the Under Secretary of Defense for Research and Engineering. This research used resources of the National Energy Research Scientific Computing Center (NERSC), a Department of Energy Office of Science User Facility using NERSC award GenAI@NERSC and MIT SuperCloud cluster. H.T. acknowledges support from the Mathworks Engineering Fellowship. The authors also thank Xiaochen Du, Juno Nam, Ethan Suresh, and Jinyu Zhang for detailed feedback on the manuscript. 

\bibliography{bibliography}

@ARTICLE{LiKELQMDY11,
	author	    = {Li, J and Kushima, A and Eapen, J and Lin, X and Qian,
	  XF and Mauro, JC and Diep, P and Yip, S},
	title	    = {Computing the Viscosity of Supercooled Liquids: Markov
	  Network Model},
	journal     = {PLoS One},
	year	    = {2011},
	volume	    = {6},
	pages	    = {e17909}
}

@ARTICLE{MishinAL10,
	author	    = {Mishin, Y and Asta, M and Li, J},
	title	    = {Atomistic modeling of interfaces and their impact on
	  microstructure and properties},
	journal     = {Acta Mater.},
	year	    = {2010},
	volume	    = {58},
	pages	    = {1117-1151}
}

@ARTICLE{AbdelhafizTZWRHZL23,
        author      = {Abdelhafiz, A and Tanvir, ANM and Zeng, MX and Wang,
          BM and Ren, ZC and Harutyunyan, AR and Zhang, YL and Li, J},
        title       = {Pulsed Light Synthesis of High Entropy Nanocatalysts
          with Enhanced Catalytic Activity and Prolonged Stability for Oxygen
          Evolution Reaction},
        journal     = {Adv. Sci.},
        year        = {2023},
        volume      = {10},
        pages       = {2300426}
}

@article{nam2025flow,
  title={Flow matching for accelerated simulation of atomic transport in crystalline materials},
  author={Nam, Juno and Liu, Sulin and Winter, Gavin and Jun, KyuJung and Yang, Soojung and G{\'o}mez-Bombarelli, Rafael},
  journal={Nature Machine Intelligence},
  pages={1--11},
  year={2025},
  publisher={Nature Publishing Group UK London}
}

@article{voter1997hyperdynamics,
  title={Hyperdynamics: Accelerated molecular dynamics of infrequent events},
  author={Voter, Arthur F},
  journal={Physical Review Letters},
  volume={78},
  number={20},
  pages={3908},
  year={1997},
  publisher={APS}
}

@article{hara2010adaptive,
  title={Adaptive strain-boost hyperdynamics simulations of stress-driven atomic processes},
  author={Hara, Shotaro and Li, Ju},
  journal={Physical Review B—Condensed Matter and Materials Physics},
  volume={82},
  number={18},
  pages={184114},
  year={2010},
  publisher={APS}
}

@book{bertsekas1996neuro,
  title={Neuro-dynamic programming},
  author={Bertsekas, Dimitri and Tsitsiklis, John N},
  year={1996},
  publisher={Athena Scientific}
}

@article{ding2018tunable,
  title={Tunable stacking fault energies by tailoring local chemical order in CrCoNi medium-entropy alloys},
  author={Ding, Jun and Yu, Qin and Asta, Mark and Ritchie, Robert O},
  journal={Proceedings of the National Academy of Sciences},
  volume={115},
  number={36},
  pages={8919--8924},
  year={2018},
  publisher={National Acad Sciences}
}

@article{shlesinger2006search,
  title={Search research},
  author={Shlesinger, Michael F},
  journal={Nature},
  volume={443},
  number={7109},
  pages={281--282},
  year={2006},
  publisher={Nature Publishing Group UK London}
}

@article{walsh2021magnetically,
  title={Magnetically driven short-range order can explain anomalous measurements in CrCoNi},
  author={Walsh, Flynn and Asta, Mark and Ritchie, Robert O},
  journal={Proceedings of the National Academy of Sciences},
  volume={118},
  number={13},
  pages={e2020540118},
  year={2021},
  publisher={National Acad Sciences}
}

@article{truhlar1996current,
  title={Current status of transition-state theory},
  author={Truhlar, Donald G and Garrett, Bruce C and Klippenstein, Stephen J},
  journal={The Journal of physical chemistry},
  volume={100},
  number={31},
  pages={12771--12800},
  year={1996},
  publisher={ACS Publications}
}

@article{ferrari2023simulating,
  title={Simulating short-range order in compositionally complex materials},
  author={Ferrari, Alberto and K{\"o}rmann, Fritz and Asta, Mark and Neugebauer, J{\"o}rg},
  journal={Nature Computational Science},
  volume={3},
  number={3},
  pages={221--229},
  year={2023},
  publisher={Nature Publishing Group US New York}
}

@article{otto2013influences,
  title={The influences of temperature and microstructure on the tensile properties of a CoCrFeMnNi high-entropy alloy},
  author={Otto, Frederik and Dlouh{\`y}, A and Somsen, Ch and Bei, Hongbin and Eggeler, Gunther and George, Easo P},
  journal={Acta Materialia},
  volume={61},
  number={15},
  pages={5743--5755},
  year={2013},
  publisher={Elsevier}
}

@article{gludovatz2014fracture,
  title={A fracture-resistant high-entropy alloy for cryogenic applications},
  author={Gludovatz, Bernd and Hohenwarter, Anton and Catoor, Dhiraj and Chang, Edwin H and George, Easo P and Ritchie, Robert O},
  journal={Science},
  volume={345},
  number={6201},
  pages={1153--1158},
  year={2014},
  publisher={American Association for the Advancement of Science}
}

@book{frenkel2023understanding,
  title={Understanding molecular simulation: from algorithms to applications},
  author={Frenkel, Daan and Smit, Berend},
  year={2023},
  publisher={Elsevier}
}

@article{wagner2022effects,
  title={Effects of Cr/Ni ratio on physical properties of Cr-Mn-Fe-Co-Ni high-entropy alloys},
  author={Wagner, Christian and Ferrari, Alberto and Schreuer, J{\"u}rgen and Couzini{\'e}, Jean-Philippe and Ikeda, Yuji and K{\"o}rmann, Fritz and Eggeler, Gunther and George, Easo P and Laplanche, Guillaume},
  journal={Acta Materialia},
  volume={227},
  pages={117693},
  year={2022},
  publisher={Elsevier}
}

@article{yao2018carbothermal,
  title={Carbothermal shock synthesis of high-entropy-alloy nanoparticles},
  author={Yao, Yonggang and Huang, Zhennan and Xie, Pengfei and Lacey, Steven D and Jacob, Rohit Jiji and Xie, Hua and Chen, Fengjuan and Nie, Anmin and Pu, Tiancheng and Rehwoldt, Miles and others},
  journal={Science},
  volume={359},
  number={6383},
  pages={1489--1494},
  year={2018},
  publisher={American Association for the Advancement of Science}
}

@article{walsh2023reconsidering,
  title={Reconsidering short-range order in complex concentrated alloys},
  author={Walsh, Flynn and Abu-Odeh, Anas and Asta, Mark},
  journal={MRS Bulletin},
  volume={48},
  number={7},
  pages={753--761},
  year={2023},
  publisher={Springer}
}

@article{tang2024reinforcement,
  title={Reinforcement Learning-Guided Long-Timescale Simulation of Hydrogen Transport in Metals},
  author={Tang, Hao and Li, Boning and Song, Yixuan and Liu, Mengren and Xu, Haowei and Wang, Guoqing and Chung, Heejung and Li, Ju},
  journal={Advanced Science},
  volume={11},
  number={5},
  pages={2304122},
  year={2024},
  publisher={Wiley Online Library}
}

@article{cowley1965short,
  title={Short-range order and long-range order parameters},
  author={Cowley, JM},
  journal={Physical Review},
  volume={138},
  number={5A},
  pages={A1384},
  year={1965},
  publisher={APS}
}

@article{shen2021kinetic,
  title={Kinetic Monte Carlo simulation framework for chemical short-range order formation kinetics in a multi-principal-element alloy},
  author={Shen, Zeqi and Du, Jun-Ping and Shinzato, Shuhei and Sato, Yuji and Yu, Peijun and Ogata, Shigenobu},
  journal={Computational Materials Science},
  volume={198},
  pages={110670},
  year={2021},
  publisher={Elsevier}
}

@article{zhang2020short,
  title={Short-range order and its impact on the CrCoNi medium-entropy alloy},
  author={Zhang, Ruopeng and Zhao, Shiteng and Ding, Jun and Chong, Yan and Jia, Tao and Ophus, Colin and Asta, Mark and Ritchie, Robert O and Minor, Andrew M},
  journal={Nature},
  volume={581},
  number={7808},
  pages={283--287},
  year={2020},
  publisher={Nature Publishing Group UK London}
}

@article{uberuaga2020computational,
  title={Computational Methods for Long-Timescale Atomistic Simulations},
  author={Uberuaga, Blas Pedro and Perez, Danny},
  journal={Handbook of Materials Modeling: Methods: Theory and Modeling},
  pages={683--688},
  year={2020},
  publisher={Springer}
}

@article{li2011diffusive,
  title={Diffusive molecular dynamics and its application to nanoindentation and sintering},
  author={Li, Ju and Sarkar, Sanket and Cox, William T and Lenosky, Thomas J and Bitzek, Erik and Wang, Yunzhi},
  journal={Physical Review B},
  volume={84},
  number={5},
  pages={054103},
  year={2011},
  publisher={APS}
}

@article{deng2023chgnet,
  title={CHGNet as a pretrained universal neural network potential for charge-informed atomistic modelling},
  author={Deng, Bowen and Zhong, Peichen and Jun, KyuJung and Riebesell, Janosh and Han, Kevin and Bartel, Christopher J and Ceder, Gerbrand},
  journal={Nature Machine Intelligence},
  volume={5},
  number={9},
  pages={1031--1041},
  year={2023},
  publisher={Nature Publishing Group UK London}
}

@article{takamoto2022towards,
  title={Towards universal neural network potential for material discovery applicable to arbitrary combination of 45 elements},
  author={Takamoto, So and Shinagawa, Chikashi and Motoki, Daisuke and Nakago, Kosuke and Li, Wenwen and Kurata, Iori and Watanabe, Taku and Yayama, Yoshihiro and Iriguchi, Hiroki and Asano, Yusuke and others},
  journal={Nature Communications},
  volume={13},
  number={1},
  pages={2991},
  year={2022},
  publisher={Nature Publishing Group UK London}
}

@article{henkelman2018long,
  title={Long-timescale simulations: Challenges, pitfalls, best practices, for development and applications},
  author={Henkelman, Graeme and J{\'o}nsson, Hannes and Leli{\`e}vre, Tony and Mousseau, Normand and Voter, Arthur F},
  journal={Handbook of Materials Modeling, Andreoni W., Yip S., Eds.(Springer, 2020)},
  pages={1--10},
  year={2018}
}

@incollection{voter2007introduction,
  title={Introduction to the kinetic Monte Carlo method},
  author={Voter, Arthur F},
  booktitle={Radiation effects in solids},
  pages={1--23},
  year={2007},
  publisher={Springer}
}

@article{voter1998parallel,
  title={Parallel replica method for dynamics of infrequent events},
  author={Voter, Arthur F},
  journal={Physical Review B},
  volume={57},
  number={22},
  pages={R13985},
  year={1998},
  publisher={APS}
}

@article{sarkar2012finding,
  title={Finding activation pathway of coupled displacive-diffusional defect processes in atomistics: Dislocation climb in fcc copper},
  author={Sarkar, Sanket and Li, Ju and Cox, William T and Bitzek, Erik and Lenosky, Thomas J and Wang, Yunzhi},
  journal={Physical Review B—Condensed Matter and Materials Physics},
  volume={86},
  number={1},
  pages={014115},
  year={2012},
  publisher={APS}
}

@article{sheriff2024quantifying,
  title={Quantifying chemical short-range order in metallic alloys},
  author={Sheriff, Killian and Cao, Yifan and Smidt, Tess and Freitas, Rodrigo},
  journal={Proceedings of the National Academy of Sciences},
  volume={121},
  number={25},
  pages={e2322962121},
  year={2024},
  publisher={National Acad Sciences}
}

@article{ziehl2023detection,
  title={Detection and impact of short-range order in medium/high-entropy alloys},
  author={Ziehl, Tyler Joe and Morris, David and Zhang, Peng},
  journal={Iscience},
  volume={26},
  number={3},
  year={2023},
  publisher={Elsevier}
}

@article{du2022chemical,
  title={Chemical domain structure and its formation kinetics in CrCoNi medium-entropy alloy},
  author={Du, Jun-Ping and Yu, Peijun and Shinzato, Shuhei and Meng, Fan-Shun and Sato, Yuji and Li, Yangen and Fan, Yiwen and Ogata, Shigenobu},
  journal={Acta Materialia},
  volume={240},
  pages={118314},
  year={2022},
  publisher={Elsevier}
}

@article{li2024tunable,
  title={Tunable interstitial and vacancy diffusivity by chemical ordering control in CrCoNi medium-entropy alloy},
  author={Li, Yangen and Du, Jun-Ping and Shinzato, Shuhei and Ogata, Shigenobu},
  journal={npj Computational Materials},
  volume={10},
  number={1},
  pages={134},
  year={2024},
  publisher={Nature Publishing Group UK London}
}

@article{grambow2020deep,
  title={Deep learning of activation energies},
  author={Grambow, Colin A and Pattanaik, Lagnajit and Green, William H},
  journal={The journal of physical chemistry letters},
  volume={11},
  number={8},
  pages={2992--2997},
  year={2020},
  publisher={ACS Publications}
}

@article{coley2019graph,
  title={A graph-convolutional neural network model for the prediction of chemical reactivity},
  author={Coley, Connor W and Jin, Wengong and Rogers, Luke and Jamison, Timothy F and Jaakkola, Tommi S and Green, William H and Barzilay, Regina and Jensen, Klavs F},
  journal={Chemical science},
  volume={10},
  number={2},
  pages={370--377},
  year={2019},
  publisher={Royal Society of Chemistry}
}

@article{cao2024capturing,
  title={Capturing short-range order in high-entropy alloys with machine learning potentials},
  author={Cao, Yifan and Sheriff, Killian and Freitas, Rodrigo},
  journal={arXiv preprint arXiv:2401.06622},
  year={2024}
}

@article{batatia2023foundation,
  title={A foundation model for atomistic materials chemistry},
  author={Batatia, Ilyes and Benner, Philipp and Chiang, Yuan and Elena, Alin M and Kov{\'a}cs, D{\'a}vid P and Riebesell, Janosh and Advincula, Xavier R and Asta, Mark and Baldwin, William J and Bernstein, Noam and others},
  journal={arXiv preprint arXiv:2401.00096},
  year={2023}
}

@inproceedings{schutt2021equivariant,
  title={Equivariant message passing for the prediction of tensorial properties and molecular spectra},
  author={Sch{\"u}tt, Kristof and Unke, Oliver and Gastegger, Michael},
  booktitle={International Conference on Machine Learning},
  pages={9377--9388},
  year={2021},
  organization={PMLR}
}

@article{mcinnes2018umap,
  title={Umap: Uniform manifold approximation and projection for dimension reduction},
  author={McInnes, Leland and Healy, John and Melville, James},
  journal={arXiv preprint arXiv:1802.03426},
  year={2018}
}

@article{shannon1948mathematical,
  title={A mathematical theory of communication},
  author={Shannon, Claude Elwood},
  journal={The Bell system technical journal},
  volume={27},
  number={3},
  pages={379--423},
  year={1948},
  publisher={Nokia Bell Labs}
}

@incollection{jonsson1998nudged,
  title={Nudged elastic band method for finding minimum energy paths of transitions},
  author={J{\'o}nsson, Hannes and Mills, Greg and Jacobsen, Karsten W},
  booktitle={Classical and quantum dynamics in condensed phase simulations},
  pages={385--404},
  year={1998},
  publisher={World Scientific}
}

@article{henkelman2000improved,
  title={Improved tangent estimate in the nudged elastic band method for finding minimum energy paths and saddle points},
  author={Henkelman, Graeme and J{\'o}nsson, Hannes},
  journal={The Journal of chemical physics},
  volume={113},
  number={22},
  pages={9978--9985},
  year={2000},
  publisher={American Institute of Physics}
}

@article{larsen2017atomic,
  title={The atomic simulation environment—a Python library for working with atoms},
  author={Larsen, Ask Hjorth and Mortensen, Jens J{\o}rgen and Blomqvist, Jakob and Castelli, Ivano E and Christensen, Rune and Du{\l}ak, Marcin and Friis, Jesper and Groves, Michael N and Hammer, Bj{\o}rk and Hargus, Cory and others},
  journal={Journal of Physics: Condensed Matter},
  volume={29},
  number={27},
  pages={273002},
  year={2017},
  publisher={IOP Publishing}
}

@article{li2023evolution,
  title={Evolution of short-range order and its effects on the plastic deformation behavior of single crystals of the equiatomic Cr-Co-Ni medium-entropy alloy},
  author={Li, Le and Chen, Zhenghao and Kuroiwa, Shogo and Ito, Mitsuhiro and Yuge, Koretaka and Kishida, Kyosuke and Tanimoto, Hisanori and Yu, Yue and Inui, Haruyuki and George, Easo P},
  journal={Acta Materialia},
  volume={243},
  pages={118537},
  year={2023},
  publisher={Elsevier}
}

@article{he2024quantifying,
  title={Quantifying short-range order using atom probe tomography},
  author={He, Mengwei and Davids, William J and Breen, Andrew J and Ringer, Simon P},
  journal={Nature Materials},
  volume={23},
  number={9},
  pages={1200--1207},
  year={2024},
  publisher={Nature Publishing Group UK London}
}

@article{nakajima1997discovery,
  title={The discovery and acceptance of the Kirkendall Effect: The result of a short research career},
  author={Nakajima, Hideo},
  journal={JoM},
  volume={49},
  number={6},
  pages={15--19},
  year={1997},
  publisher={Springer}
}

@article{zhao2018effect,
  title={Effect of d electrons on defect properties in equiatomic NiCoCr and NiCoFeCr concentrated solid solution alloys},
  author={Zhao, Shijun and Egami, Takeshi and Stocks, G Malcolm and Zhang, Yanwen},
  journal={Physical Review Materials},
  volume={2},
  number={1},
  pages={013602},
  year={2018},
  publisher={APS}
}

@article{metropolis1953equation,
  title={Equation of state calculations by fast computing machines},
  author={Metropolis, Nicholas and Rosenbluth, Arianna W and Rosenbluth, Marshall N and Teller, Augusta H and Teller, Edward},
  journal={The journal of chemical physics},
  volume={21},
  number={6},
  pages={1087--1092},
  year={1953},
  publisher={American Institute of Physics}
}

@article{li2024vacancy,
  title={Vacancy formation free energy in concentrated alloys: Equilibrium vs. random sampling},
  author={Li, Kangming and Schuler, Thomas and Fu, Chu-Chun and Nastar, Maylise},
  journal={Acta Materialia},
  volume={281},
  pages={120355},
  year={2024},
  publisher={Elsevier}
}

@article{pfeiler1988investigation,
  title={Investigation of short-range order by electrical resistivity measurement},
  author={Pfeiler, Wolfgang},
  journal={Acta Metallurgica},
  volume={36},
  number={9},
  pages={2417--2434},
  year={1988},
  publisher={Elsevier}
}

@article{rossiter1971dependence,
  title={The dependence of electrical resistivity on short-range order},
  author={Rossiter, PL and Wells, P},
  journal={Journal of Physics C: Solid State Physics},
  volume={4},
  number={3},
  pages={354},
  year={1971},
  publisher={IOP Publishing}
}

@book{rossiter1991electrical,
  title={The electrical resistivity of metals and alloys},
  author={Rossiter, Paul L},
  volume={6},
  year={1991},
  publisher={Cambridge university press}
}

\clearpage
\pagebreak
\setcounter{section}{0}
\setcounter{equation}{0}
\setcounter{figure}{0}
\setcounter{table}{0}
\setcounter{page}{1}
\makeatletter
\renewcommand{\theequation}{S\arabic{equation}}
\renewcommand{\thesection}{S\arabic{section}}
\renewcommand{\thefigure}{S\arabic{figure}}

\section{Supplementary Information }

\subsection{General expression of MFPT}
As defined in the Eq. 4, MFPT from $s_0$ is given by the probability-weighted average of first-passage times over all trajectories $\mathcal{T} \in \text{FP}(s_0)$.
The total probability $P(\mathcal T)$ is the product of the individual transition probabilities, where each transition probability $P(s\to s')$ from $s$ to $s'$ is defined as: 
\begin{equation}
    P(s\to s') \equiv \frac{\Gamma (a_{s\to s'})}{\sum_{a'\in \mathcal A_s} \Gamma (a')}
\end{equation}
The average simulation time for each trajectory, $t_{\mathcal{T}}$, is given by:  
\begin{equation}
    t_{\mathcal{T}} = \sum_{k=0}^{K-1} t_{\rm Residence}(s_k).
\end{equation}  

To justify this summation form, we derive the conditional average time associated with a given action, corresponding to a transition from the current state to a specific next state.
The joint probability distribution of action and time is as follow:
\begin{equation}
P(a, t) d t=e^{-\left(\sum_{a^{\prime} \in \mathcal A_s} \Gamma (a^{\prime})\right) t} \Gamma (a) d t
\end{equation}
as the incident $(a,t)$ means “action $a$ happens in the time interval $[t,t+dt]$, and no action happens before that time interval”. The probability that no action happens before that time interval is $e^{-\left(\sum_{a^{\prime} \in \mathcal A_s} \Gamma (a^{\prime})\right) t}$, and the probability that action $a$ happens in the time interval $[t,t+dt]$ equals $\Gamma (a) dt$. 
The conditional probability is then derived as
\begin{equation}
\begin{aligned}
P(t\mid a) &=\frac{P(a, t)}{P(a)}=\frac{e^{-\left(\sum_{a^{\prime} \in \mathcal A_s} \Gamma (a^{\prime})\right) t} \Gamma (a)}{\Gamma (a) / \sum_{a^{\prime} \in \mathcal A_s} \Gamma (a^{\prime})}\\
&=e^{-\left(\sum_{a^{\prime} \in A_s} \Gamma (a^{\prime})\right) t} \sum_{a^{\prime} \in \mathcal A_s} \Gamma (a^{\prime})
\end{aligned}
\end{equation}
Eventually, the conditional average time is
\begin{equation}
\mathbb{E}[t \mid a]=\int_0^{\infty} P(t\mid a) t d t=\frac{1}{\sum_{a^{\prime} \in A_s} \Gamma_{a^{\prime}}}=t_{\text {Residence }}
\end{equation}
Counterintuitively, the conditional average time given an action is the same as the average time itself. Therefore, the conditional average time given the trajectory is also obtained from the residence times summed over all timesteps in the trajectory.

\subsection{Reaction model performance}\label{model_performance}
To assess the predictive accuracy of our reaction model, we first trained it on the same dataset of hydrogen diffusion in CrCoNi used in prior work~\cite{tang2024reinforcement}. As shown in Fig.~\ref{fig:h_accuracy}, the current model demonstrates improved performance, achieving root mean square error (RMSE) values of 0.024 eV for $E_{\rm A}$ and 0.05 for $\ln(\nu_{\rm A})$, compared to the previous model, which had RMSE values of 0.037 eV and 0.12, respectively. As shown in Fig. 2b and S6-8, we further evaluated the predictive accuracy of the two models in two different test sets: an in-distribution test set consisting of 108, 256, and 500 atoms, and an out-of-distribution set consisting of 864 atoms. The in-distribution test set contains configurations similar to the training data, while the out-of-distribution test set represents a significantly larger system, challenging the model performance of the scalability to larger systems. In both scenarios, the GNN-based reaction model outperformed the previous method in predicting both $E_{\rm A}$ and $\nu_{\rm A}$.

\subsection{DRL-accelerated atomistic simulation}
We compared the temperature-dependent evolution of SRO using three simulation approaches: transition-kinetics sampling (TKS, equivalent to kinetic Monte Carlo), vacancy-constrained Metropolis Monte Carlo (MMC$^{\rm vac}$), and DRL-LSS (Fig.~\ref{fig:comparison}). In all simulations, the action space was restricted to vacancy hops between first nearest-neighbor atomic sites. To minimize artifacts arising from specific initial configurations, each simulation was independently repeated five times per temperature using randomized RSS structures.
At sufficiently high temperatures (e.g., 900~K), all three methods exhibit similar convergence in both energy and SRO, indicating that thermal fluctuations dominate the sampling process. However, at lower temperatures (e.g., 300~K), TKS trajectories tend to become trapped in high-energy disordered states due to repeated forward–backward transitions—a known bottleneck in kMC-based simulations under limited thermal activation.
In contrast, DRL-LSS more effectively guides the system toward lower-energy configurations with higher degrees of SRO, demonstrating a superior ability to overcome kinetic trapping. Notably, it outperforms even MMC$^{\rm vac}$, highlighting the advantage of long-horizon planning inherent in DRL-based approaches.

\subsection{Training objective for timescape  estimator}\label{time_estimation_method}

To improve training stability, we refine the learning objective of the timescape estimator.
According to Eq. 5, in principle, when $s \notin\mathcal G$, MFPT($s$), can be updated by sampling a transition from $s$ to $s'$ as shown in Eq.~\eqref{eq:bellman_update}.
\begin{equation}
\begin{aligned}
\text{MFPT}(s) \leftarrow \text{MFPT}(s) + \\
\eta \left[\mathbbm{1}_{s \notin \mathcal G} \cdot \left(t_{\rm Residence}(s) + \text{MFPT}(s')\right) - \text{MFPT}(s)\right]
    \label{eq:bellman_update}
\end{aligned}
\end{equation}
where $\mathbbm{1}_{s\notin \mathcal G}$ equals one when $s$ is not in $\mathcal G$ and zero otherwise, and $\eta$ is a learning rate parameter. However, such iteration does not necessarily guarantee convergence. According to Proposition 4.4 in Ref.~\cite{bertsekas1996neuro}, the convergence of the iteration requires the max norm contraction of the update to MFPT, which can be realized with a discount factor $\gamma(s)$ on the right-hand side. 
Thus, we utilized a modified MFPT$_\tau$ by introducing the $\gamma(s)$ to stabilize the convergence of MFPT training. 
The MFPT$_\tau$ is defined as follows:
\begin{equation}
    \text {MFPT}_\tau(s)
   \equiv f_{\tau}^{-1}\left(\sum_{\mathcal T\in \text{FP}(s)} P(\mathcal T) f_{\tau}(t_{\mathcal T})\right)
\end{equation}
with a kernel function $f_\tau (x) = \int_0^x e^{-t/\tau}dt=\tau(1-e^{-x/\tau})$. When $\tau$ goes to infinity, $\text{MFPT}_\tau$ converges to MFPT. One can consider $\text{MFPT}_\tau$ as another way to average the trajectory time, namely $f_\tau^{-1}(\mathbb{E}_{\mathcal T} f_\tau (t_{\mathcal T}))$.
Then, the Bellman-type equation 5 can be rewritten as:
\begin{equation}
\begin{aligned}
    f_\tau(\text{MFPT}_{\tau} (s)) =& 
   \mathbbm{1}_{s\notin \mathcal G} \{f_\tau (t_{\rm Residence}(s)) \\
    &+ \gamma(s)\mathbb E_{s'\sim P(s\to s')} f_\tau(\text{MFPT}_{\tau} (s'))\}.
    \label{eq:bellman_rewritten}
\end{aligned}
\end{equation}
In this form, a discount factor $\gamma (s) = e^{-t_{\rm Residence}(s)/\tau}<1$ is introduced, making the update function a max-norm contraction. The value of $\tau$ should be a few orders of magnitude larger than the $t_{\rm Residence}$ of simulated processes, so that $\gamma (s)$ is neither too close to nor too far from 1.

The timescape estimator, $t_\theta(s)$, is then used to fit $ f_\tau(\text{MFPT} (s))$. The equation can be written as the following optimization problem of the loss function:
\begin{equation}
\begin{aligned}
    \mathcal{L}(\theta_1, \theta_2) &= \mathbb{E}_{s\sim P(s)} \mathbb{E}_{s'\sim P(s\to s')} \Bigl[ 
    \mathbbm{1}_{s\notin \mathcal{G}} \bigl( t_{\theta_1}(s) \\&
    - f_\tau (t_{\rm Residence}(s)) - \gamma(s)  t_{\theta_2} (s') \bigr)^2 \\&
    \quad +\mathbbm{1}_{s\in \mathcal{G}} t_{\theta_1}(s)^2 \Bigr],\\
    \theta_2 &= \arg\min_{\theta_1} \mathcal{L}(\theta_1 ,\theta_2).
    \label{eq:scf}
\end{aligned}
\end{equation}
where $P(s)$ is the probability distribution function of the states sampled by kinetics simulations. If $t_\theta$ can take exactly the functional form of $ f_\tau(\text{MFPT} (s))$, the self-consistently minimized $\theta$ gives an $t_\theta$ that satisfies Eq.~\eqref{eq:bellman_rewritten}. 
In practice, we adopt TD learning and update $\theta$ using the gradient descent method:
\begin{equation}
\begin{aligned}
    \mathcal{L}_{\rm data}(\theta_1, \theta_2) &= \sum_{(s,s')} \Bigl[ 
    \omega\mathbbm{1}_{s\notin \mathcal{G}} \bigl( t_{\theta_1} (s) \\&
    - f_\tau (t_{\rm Residence}(s)) - \gamma(s)  t_{\theta_2} (s') \bigr)^2 \\&
    \quad + \mathbbm{1}_{s\in \mathcal{G}} t_{\theta_1}(s)^2 \Bigr],\\
    \theta &\leftarrow \theta - \eta \nabla_{\theta_1} \mathcal{L}_{\rm data}(\theta_1, \theta_2) \Big|_{\theta_1 = \theta_2 = \theta}.
    \label{eq:train}
\end{aligned}
\end{equation}
where the summation goes through the current state-next state pairs $(s, s')$ in the sampled trajectories in the training dataset. The $\omega$ is the weight coefficient. 
To capture the temperature and vacancy concentration dependence of MFPT more efficiently, the model needs to distinguish the change of these two variables. 
We assume that the kinetic relaxation timescale is inversely proportional to the vacancy concentration and follows the Arrhenius form of temperature dependence. 
Therefore, a scaler with the following form is designed as $\text{scaler}(X_{\rm vac}, T) = \frac{X_{\rm vac}^0}{X_{\rm vac}}e^{-\frac{Q_{\rm A} }{k_{\rm B}}(\frac{1}{T} -\frac{1}{T_0})}$.
In this work, we adopt $\frac{1}{256}$ and 500 K for $X_{\rm vac}^0$ and $T_0$, respectively. 

Then we estimate the time when $(X_{\rm vac}, T)\neq (X_{\rm vac}^0, T_0)$ as $t_\theta (s) = \text{scaler}(X_{\rm vac}, T)\cdot \tilde{t}_\theta (s)$. Accordingly, $\tau_{X_{\rm vac}, T}$ and $\gamma_{X_{\rm vac}, T}(s)$ are scaled with the scaler function as:
\begin{equation}
\notag
\begin{aligned}
    \tau_{X_{\rm vac}, T} = \tau_0\times  \text{scaler}(X_{\rm vac}, T), \ \ \ \\\gamma_{X_{\rm vac}, T}(s) = e^{-t_{\rm Residence}(s)/\tau_{X_{\rm vac}, T} }.
\end{aligned}
\end{equation}
The loss function is then modified to effectively account for training data for all parameters with uniform weights:
\begin{equation}
\begin{aligned}
    \mathcal L_{\rm data}(\theta_1, \theta_2) =& \sum_{(s,s',X_{\rm vac},T)} \Bigl[ \omega \mathbbm{1}_{s\notin \mathcal G}\bigl(\tilde{t}_{\theta_1}\\
    &-\frac{f_{\tau_{X_{\rm vac}, T}} (t_{\rm Residence}(s))}{\text{scaler}(X_{\rm vac},T)} \\&- \gamma_{X_{\rm vac}, T}(s) \tilde{t}_{\theta_2} (s')\bigr)^2 \\&
    +\omega_{\mathcal G} \mathbbm{1}_{s\in \mathcal G}\tilde{t}_{\theta_1} (s)^2\Bigl].
    \end{aligned}
    \label{eq:time_loss_full}
\end{equation}

In multitask learning, the timescape estimator outputs a two-dimensional vector, $(y, z)$, where $y$ is used for regression to predict time, and $z$ is used for binary classification to produce logits that determine whether $s$ belongs to $\mathcal{G}$. The final model output, $\tilde{t}_\theta$, is then computed as:
\begin{equation}
\notag
\tilde{t}_\theta = 
\begin{cases}
0, & \text{if } \sigma(z) \geq 0.5 \\
y, & \text{otherwise},
\end{cases}
\label{eq:time_estimator}
\end{equation}
where \( \sigma(z) = \frac{1}{1 + e^{-z}} \) is the sigmoid function applied to the logits. A classifier $z_\theta (s)$ is trained to distinguish whether the state $s$ is in $\mathcal G$ or not, as shown in Eq.~\eqref{eq:time_estimator}. The loss function is then modified to effectively account for training data for all parameters with uniform weights:
\begin{equation}
\begin{aligned}
    \mathcal L_{\rm data}(\theta_1, \theta_2) =& \sum_{(s,s',X_{\rm vac},T)} \Bigl[ \omega_{\rm Cls}\mathcal{L}_{\text{BCE}}(\mathbbm{1}_{s\in \mathcal G}, z_{\theta_1} (s)) \\&+ \mathbbm{1}_{s\in \mathcal G}\tilde{t}_{\theta_1} (s)^2\\ 
    &+\omega \mathbbm{1}_{s\notin \mathcal G}\bigl(\tilde{t}_{\theta_1}-\frac{f_{\tau_{X_{\rm vac}, T}} (t_{\rm Residence}(s))}{\text{scaler}(X_{\rm vac},T)} \\
    &- \gamma_{X_{\rm vac}, T}(s) \tilde{t}_{\theta_2} (s')\bigr)^2 \Bigl].
    \end{aligned}
    \label{eq:time_loss_full}
\end{equation}
 $\mathcal{L}_{\text{BCE}}$ represents the binary cross-entropy loss~\cite{shannon1948mathematical}:

\begin{equation}
\notag
\mathcal{L}_{\text{BCE}}(\mathbbm{1}_{s\in \mathcal G},z) = - \mathbbm{1}_{s\in \mathcal G} \log\sigma(z) - (1-\mathbbm{1}_{s\in \mathcal G}) \log\bigl(1 - \sigma(z)\bigr).
\end{equation}
The converged parameter $\theta$ still gives a solution to the self-consistent equation Eq.~\eqref{eq:bellman_rewritten}. 

\subsection{Hyperparameters for timescape estimator training}
To determine the hyperparameters for training the timescape estimator $\text{MFPT}_\theta$, we used a down-sampled dataset consisting of 1,000 data points. 
The activation energy $Q_{\rm A}$ was first obtained by fitting the temperature dependence of the residence time $t_{\mathrm{Residence}}$ in the training data (see Fig.~S9), serving as the temperature scaling factor.
Based on this, we computed the discount factor $\gamma(s)$ and determined $\tau_0$ accordingly. 
In this work, we use $Q_{\rm A} = -0.935$~eV and set $\tau_0 = 2000~\mu\mathrm{s}$ (Fig. ~\ref{fig:discount_factor}).

Figure~S13 presents receiver operating characteristic (ROC) curves on the down-sampled test set, comparing timescape estimators trained with different loss functions.
A true positive is defined as the correct prediction that a state $s$ does not belong to $\mathcal{G}$. 
The area under the curve (AUC) quantifies binary classification performance across varying thresholds, with higher AUC indicating better predictive accuracy. 
Models trained with larger values of the weight coefficient $\omega$ exhibit improved performance, both in terms of AUC and Pearson's correlation coefficient $r$.
Incorporating a classification loss further enhances model accuracy, with optimal performance achieved at $\omega_{\mathrm{Cls}} = 0.2$. 
Based on these, we use $\omega = 5.0$ and $\omega_{\mathrm{Cls}} = 0.2$ in all subsequent training (Fig. ~\ref{fig:t_model_evaluation}).

        \subsection{Short-range order and electrical resistivity}
The electrical resistivity in alloys has been extensively studied in the previous literature~\cite{rossiter1991electrical}. Multiple theoretical modeling and experimental examination provide a conclusion that the electric resistivity of alloys should depend on the short-range order linearly~\cite{pfeiler1988investigation}. The residue electric resistivity of an alloy can be expressed with the relaxation time $\tau$ of electron scattering:
\begin{equation}
    \rho = \frac{m}{e^2n\tau},
\end{equation}
where $m$, $e$, and $n$ are the effective mass, charge, and density of carrier in the alloy. In principle, the relaxation time depends on multiple scattering channels:
\begin{equation}
    \frac{1}{\tau} = \frac{1}{\tau_{\rm e-phonon}} + \frac{1}{\tau_{\rm e-defect}} + \frac{1}{\tau_{\rm e-e}} + \frac{1}{\tau_{\rm e-disorder}}
\end{equation}
It is commonly assumed that the first three terms (phonon scatterning, defect scatterning, and electron-electron scatterning) are not sensitively dependent on the SRO, so their contribution can be approximated as a constant, $\rho_0$. The last term, $\frac{1}{\tau_{\rm e-disorder}}$, originates from electron scatterning with local randomness and correlation of element occupation on neighboring lattice sites. It has been shown that the electrical resistivity from the electron-disorder scattering term depends linearly on the WC parameters from theoretical modeling~\cite{rossiter1971dependence}:
\begin{equation}
    \rho_{\rm e-disorder} \propto \sum_n c^n \alpha^n Y^n,
\end{equation}
where $c^n$, $\alpha^n$, and $Y^n$ are the number of sites, WC parameter, and theoretical constants (derived from the model) of the $n$th-nearest neighbor shell. In the experimental examination in the literature, the first term ($n=1$, the first nearest neighbor) usually dominates the electric resistivity for a number of systems~\cite{pfeiler1988investigation}. Therefore, we can make an approximation that total electric resistivity has a linear relation with the first-nearest neighbor WC parameter:
\begin{equation}
    \rho = \rho_0 + \rho_{\rm disorder} = \rho_0 + c^1Y^1 \alpha^1.
    \label{linearsro}
\end{equation}
The experimentally measured relative electric resistivity change
\begin{equation}
    \frac{\rho}{\rho_0} = 1 + \frac{c^1Y^1}{\rho_0}\alpha^1
\end{equation}
is then a linear-scale measure of the WC parameter of the SRO. As the literature theory behind Eq.~\eqref{linearsro} was originally derived for binary alloy, there can be more complication in deriving the numerical factors for the HEA cases, while the linear relation will remain unchanged. 
In this work, the overall degree of SRO $\Omega_{\alpha}$, is defined as $\Omega_{\alpha}\equiv \sqrt{\sum_{i \leq j} |\alpha^1_{ij}|^2}$. 
A more rigorous study on the relationship between the electrical resistivity and the quantified SRO for multicomponent HEAs is left for future investigation.

A previous study~\cite{li2023evolution} measured the electrical resistivity as a function of annealing temperature and time and fitted the data to a simple exponential relaxation form:
\begin{equation}
    \langle\rho(t, T)\rangle
    = \rho_{\max}(T)
      - \Delta\rho\cdot e^{-t/\beta^{\rho}(T)}.
\end{equation}
In an experimental specimen, the fluctuation $\sigma^{\rho}$ that scales as $1/\sqrt{N}$ becomes negligible as the macroscopic sample contains $N \approx N_{\rm A}$ atoms. Consequently, $\rho_{\rm max}(T)=\langle \rho^{\rm Eq}\rangle$ is just the average equilibrium electrical resistivity.

The relaxation of SRO can be also expressed as a simple exponential form in adjacency of the target states:
\begin{equation}
\begin{aligned}
    \langle{\Omega_{\alpha}}(t, T)\rangle
    = {\Omega_{\alpha}}_{\max}(T)
      - \Delta\Omega_{\alpha}\cdot e^{-t/\beta^{\rm SRO}(T)}.
\end{aligned}
\end{equation}
Unlike the resistivity case, the equilibrium fluctuation for SRO is not negligible, because our simulations are performed for a finite number of atoms $N^{\rm sim}$.
Therefore, we define the target-state manifold $\mathcal{G}$ as the hypersurface corresponding to the equilibrium-fluctuation volume around the centroid with a fluctuation radius of $2\sigma^{\rm SRO}_{q}$. 
In this definition, when $|\langle{\Omega_{\alpha}}(t, T)\rangle -\langle{\Omega_{\alpha}}^{\rm Eq}\rangle|  \le2\sigma_q^{\rm SRO}$, our predicted MFPT is set to zero, and the predicted elapsed time is plateaued. Therefore, when the system is so close to equilibrium that it enters the $2\sigma_q^{\rm SRO}$ target state, the predicted time will be underestimated compared to the actual evolution time. At the position the system first enters the target states, the $\langle{\Omega_{\alpha}}(t, T)\rangle$ reaches the following percentage of $\langle{\Omega_{\alpha}}^{\rm Eq}\rangle$:
\begin{equation}
    \frac{\langle{\Omega_{\alpha}}(t, T)\rangle}{\langle{\Omega_{\alpha}}^{\rm Eq}\rangle} = (1-\frac{2\sigma_q^{\rm SRO}}{\langle{\Omega_{\alpha}}^{\rm Eq}\rangle})\times 100\%
\end{equation}
From our numerical simulations, we estimate this value to be approximately $90\%$, meaning that a majority of the SRO formation process is outside the target states and free from the underestimation issue. Although the predicted time when the SRO formation is greater than $90\%$ would be underestimated in our method, as the data we used to fit the $t_{\rm anneal}$ vs. SRO relation are mostly from the SRO smaller than $90\%$, the deviation is expected to be small.

The distance to the $\langle\Omega_{\alpha}^{\rm Eq}\rangle$ when the system enters the boundary of the $2\sigma_q^{\rm SRO}$ target states is
\begin{equation}
    \Delta(t, T)
    =  \langle{\Omega_{\alpha}}^{\rm Eq}\rangle -  \langle{\Omega_{\alpha}}(t, T)\rangle
    = \Delta \Omega_{\alpha}\times \, e^{-t/\beta^{\rm SRO}(T)}.
\end{equation}
If we define the hit-time $t_{\rm hit}$ as the time at which the trajectory first enters $\mathcal G$, the $2\sigma^{\rm SRO}_{q}$ fluctuation scale around the plateau,
\begin{equation}
    \Delta(t_{\rm hit}, T) = 2\sigma^{\rm SRO}_{q},
\end{equation}
then
\begin{equation}
    \Delta \Omega_{\alpha} \, e^{-t_{\rm hit}(T)/\beta^{\rm SRO}(T)} = 2\sigma^{\rm SRO}_{q}.
\end{equation}
Rearranging for $\beta^{\rm SRO}(T)$ yields
\begin{equation}
    t_{\rm hit}(T)
    = \beta^{\rm SRO}(T)\ln\!\left(\Delta \Omega_{\alpha} / 2\sigma^{\rm SRO}_{q} \right).
\end{equation}
Thus, the quantity $t_{\rm hit}(T)$  can be expressed directly in terms of the threshold distance $2\sigma^{\rm SRO}_{q}$. When the actual evolution time goes beyond $t_{\rm hit}$, the predicted MFPT will no longer decrease, so using the MFPT to predict the evolution time is no longer valid after the time goes beyond $t_{\rm hit}$.

For interpolating the time–temperature–SRO map, we used a parametrized NN with a stretching exponent $\gamma(T)$ to reflect the wide range of time evolution  under diverse thermodynamic conditions. The model receives the vacancy concentration employed in our simulation cell ($X_{\rm vac}$), temperature, and log-scaled time as inputs $x = [X_{\rm vac}, T, \log({t_{\rm anneal}})]$, and evaluates the SRO using the following stretched exponential form:
\begin{equation}
\begin{aligned}
    {\Omega_{\alpha}}(t,T) = {\Omega_{\alpha}}_{\max}(T)\cdot \\
    \left[1 - \exp\left( -\left( \frac{10^{\log_{10} t_{\rm anneal} - \log_{10}{\text{scaler}(X_{\rm vac}, T)}}}{\beta(T)} \right)^{\gamma(T)} \right) \right].
\end{aligned}
\end{equation}
To ensure physical plausibility, the characteristic timescale is constrained to positive values by defining $\beta(T) = \exp(\log \beta)$. The stretching exponent is bounded within the interval $(0, 1.5)$ through a scaled sigmoid transformation,
\[
\gamma(T) = 1.5 \cdot \sigma(\gamma_{\text{raw}}), \quad \text{with } \sigma(z) = \frac{1}{1 + e^{-z}}.
\]
This parameterization enables the model to represent a wide range of relaxation behaviors, from nearly single-exponential decay $(\gamma \sim 1)$ to broadly distributed sub-exponential kinetics $(\gamma < 1)$. In our work, ${\Omega_{\alpha}}_{\max}(T)$ is defined as the target states with fluctuations of $2\sigma^{\mathrm{SRO}}_{q}$.
Each parameter—${\Omega_{\alpha}}_{\max}(T)$, $\log \beta(T)$, and the raw stretching exponent $\gamma_{\text{raw}}(T)$—is expressed as a third-order polynomial in normalized temperature:
\begin{equation}
\left[{\Omega_{\alpha}}_{\max},\, \log \beta,\, \gamma_{\text{raw}}\right] = \mathbf{W} \cdot \left[1,\; \left(\frac{T}{T_0}\right),\; \left(\frac{T}{T_0}\right)^2,\; \left(\frac{T}{T_0}\right)^3\right]^{\intercal},
\end{equation}
where $\mathbf{W} \in \mathbb{R}^{3 \times 4}$ is a learnable weight matrix optimized during training. This design enables the model to flexibly capture the variation of the $\Omega_{\alpha}$ across a broad range of temperatures and vacancy concentrations. Fitted parameters are shown in Table S2.

\begin{table*}[ht]
\centering
\caption{Calculated target short-range order parameters ($\alpha^{\rm target}_{ij}$) for CrCoNi alloy at different temperatures. The parameter $\alpha^{\rm target}_{ij}$ represents the target SRO parameter for atomic pairs of elements $i$ and $j$.}
\begin{tabular}{|c|c|c|c|c|c|c|}
\hline
\multirow{2}{*}{\textbf{Temperature (K)}} & \multicolumn{6}{c|}{\textbf{Target Warren-Cowley (WC) Parameters}} \\ \cline{2-7} 
 & $\alpha^{\rm target}_{\rm CrCr}$ & $\alpha^{\rm target}_{\rm CrCo}$ & $\alpha^{\rm target}_{\rm CoCo}$ & $\alpha^{\rm target}_{\rm CrNi}$ & $\alpha^{\rm target}_{\rm CoNi}$ & $\alpha^{\rm target}_{\rm NiNi}$ \\ \hline 
300 & 0.41 & -0.25 & -0.02  & -0.16 & 0.26  & -0.09  \\ \hline
500 & 0.35 & -0.22 & 0.04  & -0.13 & 0.18  & -0.04  \\ \hline
700 & 0.27 & -0.17 & 0.04  & -0.11 & 0.13  & -0.02  \\ \hline
900 & 0.24 & -0.13 & 0.03  & -0.11 & 0.10  & 0.01  \\ \hline
\end{tabular}
\label{table:target_sro}
\end{table*}

\begin{table*}[ht]
\centering
\caption{
Fitted polynomial coefficients \( \mathbf{W}_{ij} \) for SRO interpolation, where each row corresponds to the output parameters \( {\Omega_{\alpha}}_{\max}(T) \), \( \log \beta(T) \), and \( \gamma_{\text{raw}}(T) \), respectively. Columns represent polynomial terms of order \( k = 0, 1, 2, 3 \) in \( (T/T_0)^k \).
}
\begin{tabular}{|c|c|c|c|c|}
\hline
Parameter & $(T/T_0)^0$ & $(T/T_0)^1$ & $(T/T_0)^2$ & $(T/T_0)^3$ \\
\hline
${\Omega_{\alpha}}_{\max}(T)$ & 0.6086 & 0.5329 & -0.8298 & 0.2481  \\
\hline
$\log \beta(T)$ & -4.5878 & -2.4997 & -0.8721 & 1.2499 \\
\hline
$\gamma_{\text{raw}}(T)$ & 0.2923 & -0.3377 & -1.7914 & 1.0215 \\
\hline
\end{tabular}
\label{table:nn_weights}
\end{table*} 


\begin{figure*}[t!]
\centering
\includegraphics[width=0.3\textwidth]{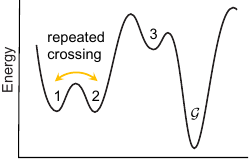}
\caption{Schematic illustration of an energy profile of a four-state system
(1, 2, 3, and $\mathcal G$).}
\label{fig:repeated_crossing}
\end{figure*}

\begin{figure*}[t]
    \centering
    \includegraphics{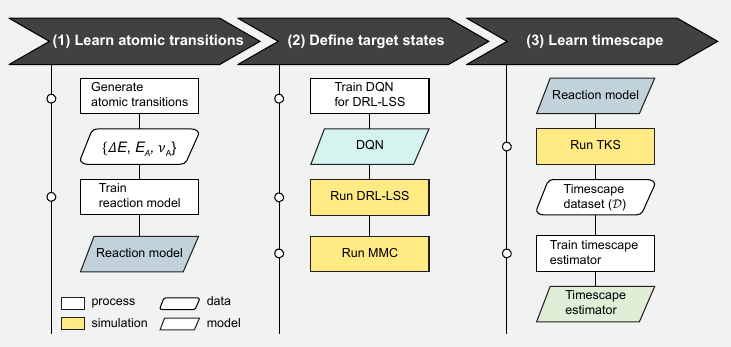}
    \caption{Computational workflow for mean first passage time (MFPT) training.
    }
    \label{fig:flow_chart}
\end{figure*}

\begin{figure*}[t]
    \centering
    \includegraphics{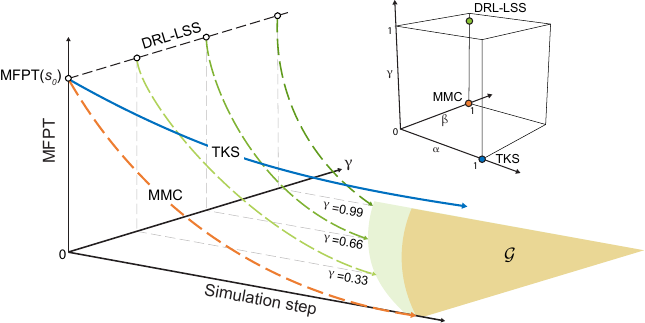}
    \caption{Schematic illustration of deep reinforcement learning (DRL)-accelerated atomistic simulations. Starting from a random solid solution ($s_0$), Metropolis Monte Carlo (MMC) sampling approaches the target states ($\mathcal{G}$). The DRL-based lower-energy state sampler (DRL-LSS) accelerates convergence toward $\mathcal{G}$ by utilizing a larger discount factor $\gamma$. The transition kinetics simulator (TKS) follows physical transition probabilities and evolves at a slower rate (shown as a solid line), whereas MMC and DRL-LSS generate non-physical transitions (shown as dashed lines). The inset figure illustrates the DRL-based simulation modes, which are governed by three hyperparameters: the kinetic and thermodynamic reward weights ($\alpha$ and $\beta$), and the discount factor $\gamma$.
    }
    \label{fig:drl_illustration}
\end{figure*}

\begin{figure*}[t]
\centering
\includegraphics{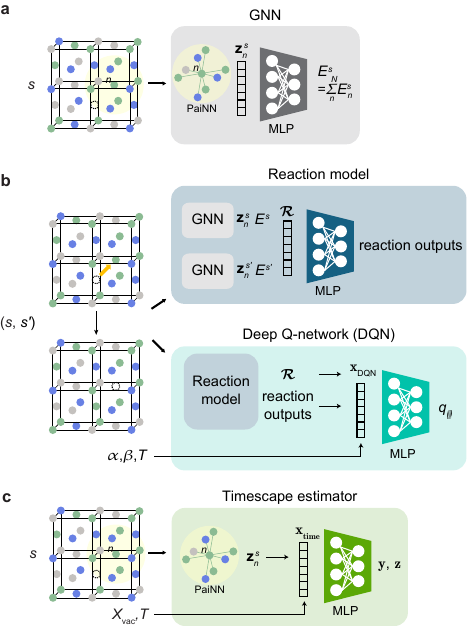}
\caption{GNN-based surrogate models. 
\textbf{a}, Schematic illustration of the GNN to encoding local atomic environments $z_n$, and predicting the total energy $E$. \textbf{b}, Learning of atomic transitions vector $\boldsymbol{\mathcal R}$, and predicts reaction outputs, which include reaction energy ($\Delta E$), activation energy ($E_{\rm A}$), and attempt frequency ($\nu_{\rm A}$). The learned $\boldsymbol{\mathcal R}$ is then used in DQN, allowing for efficient training of the DQN. MLP represents a multi-layer perceptron.
\textbf{c}, Schematic illustration of the timescape estimator architecture.
}
\label{fig:models}
\end{figure*}

\begin{figure*}[t]
\centering
\includegraphics[width=0.3\textwidth]{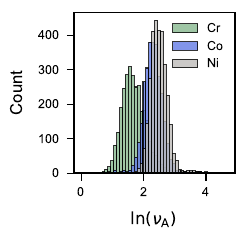}
\caption{Histogram of the attempt frequency, $\ln(\nu_{\rm A})$, for vacancy hopping events involving each atomic species. Attempt frequency values are in units of THz.}
\label{fig:distribution_nu}
\end{figure*}

\begin{figure*}[t]
\centering
\includegraphics[width=0.75\textwidth]{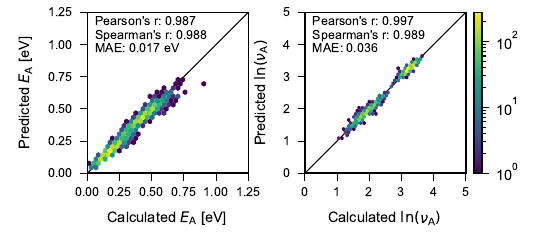}
\caption{Prediction accuracy on the activation barrier, $E_{\rm A}$ and attempt frequencies, $\ln(\nu_{\rm A})$, of hydrogen diffusion in CrCoNi for our GNN architecture.}
\label{fig:h_accuracy}
\end{figure*}

\begin{figure*}[t]
\centering
\includegraphics[width=0.75\textwidth]{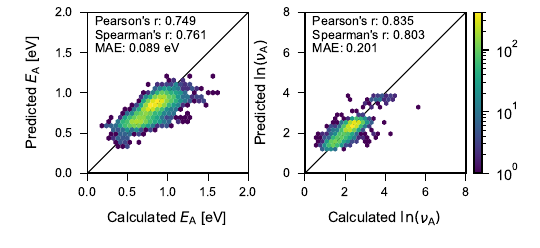}
\caption{Prediction accuracy on the energy barriers, $E_{\rm A}$, and attempt frequencies, $\ln(\nu_{\rm A})$ of the previous modified DeepPot-SE description in Ref.~\cite{tang2024reinforcement} for in-distribution test set.}
\label{fig:int_test_prev}
\end{figure*}

\begin{figure*}[t]
\centering
\includegraphics[width=0.75\textwidth]{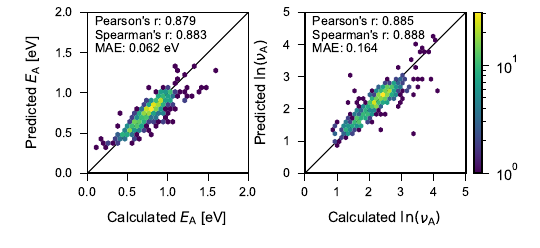}
\caption{Prediction accuracy on the energy barriers, $E_{\rm A}$, and attempt frequencies, $\ln(\nu_{\rm A})$ of the current GNN-based reaction model for out-of-distribution test set.}
\label{fig:out_test_current}
\end{figure*}

\begin{figure*}[t]
\centering
\includegraphics[width=0.75\textwidth]{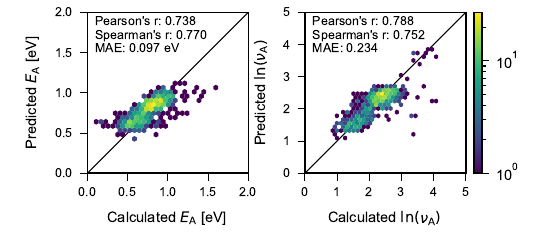}
\caption{Prediction accuracy on the energy barriers, $E_{\rm A}$, and attempt frequencies, $\ln(\nu_{\rm A})$ of the previous modified DeepPot-SE description in Ref.~\cite{tang2024reinforcement} for out-of-distribution test set.}
\label{fig:out_test_prev}
\end{figure*}

\begin{figure*}[!htb]
    \centering
    \includegraphics{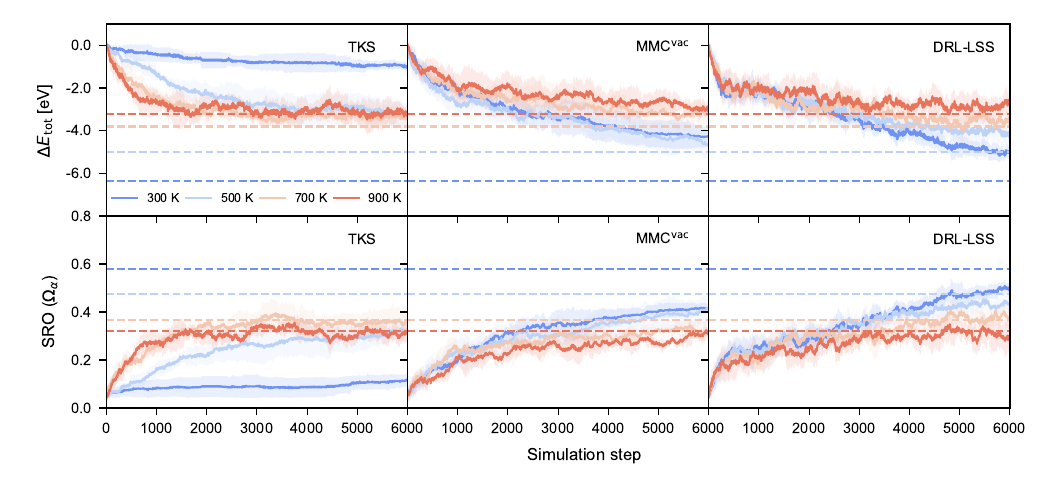}
\caption{SRO evolution in CrCoNi with 0.39~at\% vacancy concentration at different annealing temperatures ($T_{\rm anneal}$), simulated using three different modes: TKS, MCMC$^{\rm vac}$, and DRL-LSS. All modes allow only physical vacancy hopping as possible actions. MCMC$^{\rm vac}$ and DRL-LSS employ simulated annealing following the schedule $T = 1200 - \frac{1200 - T_{\rm anneal}}{4500} \times {\rm step}$ over 4500 simulation steps, followed by equilibration at a constant temperature $T_{\rm anneal}$. TKS is performed at a constant $T_{\rm anneal}$ without annealing. 
The average physical time corresponding to 6,000 TKS simulation steps at 300, 500, 700, and 900~K is $1.6\times10^2$, $2.1\times10^{-3}$, $2.0\times10^{-5}$, and $1.4\times10^{-6}$~s, respectively.}
\label{fig:comparison}
\end{figure*}

\begin{figure*}[t]
    \centering
    \includegraphics[width=0.5\textwidth]{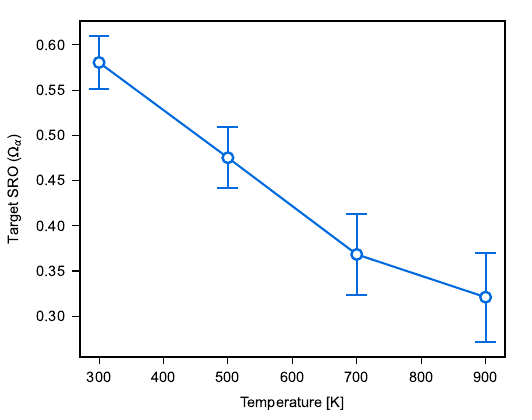}
    \caption{Thermodynamic equilibrium of SRO (${\rm SRO^{\rm Eq}}(T)$) at different temperatures. The error bars indicate the standard deviation of SRO.}
    \label{fig:thermo_limits}
\end{figure*}

\begin{figure*}[t]
\centering
\includegraphics[width=0.5\textwidth]{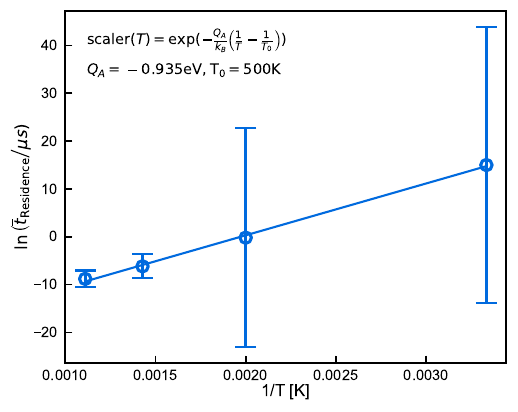}
\caption{Arrhenius-type temperature scaling of the residence time average ($\overline{t}_{\rm Residence}$) in the down-sampled training set at a given temperature, with units of $10^{-6}$ seconds. The error bars indicate the relative error, $\sigma_{t_{\rm Residence}}/\overline{t}_{\rm Residence}$.}
\label{fig:t_scaler}
\end{figure*}

\begin{figure*}[t]
\centering
\includegraphics[width=0.5\textwidth]{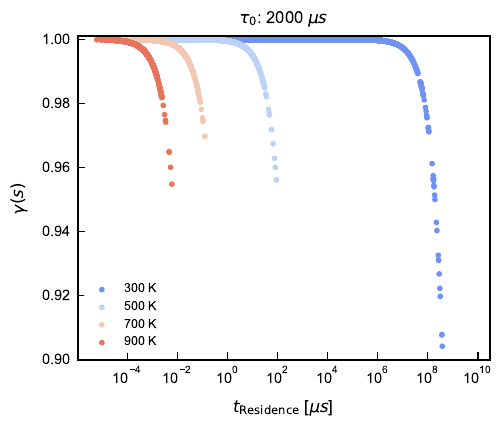}
\caption{Distribution of the discount factors $\gamma (s)$ in MFPT training with $\tau_0=2000$ $\mu$s.}
\label{fig:discount_factor}
\end{figure*}

\begin{figure*}[t]
\centering
\includegraphics{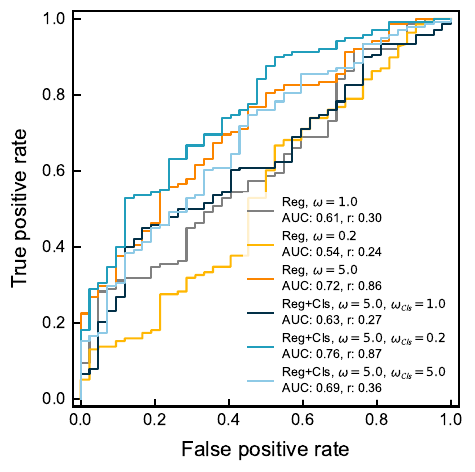}
\caption{Timescape estimator performance on down-sampled test set.  
Receiver operating characteristic (ROC) curves comparing predictive performance of the regression-only (Reg) and the combined regression-plus-classification (Reg+Cls) models with varying weight coefficients $\omega$ and $\omega_{\mathrm{Cls}}$.}
\label{fig:t_model_evaluation}
\end{figure*}

\begin{figure*}[!htb]
\centering
\includegraphics[width=0.6\textwidth]{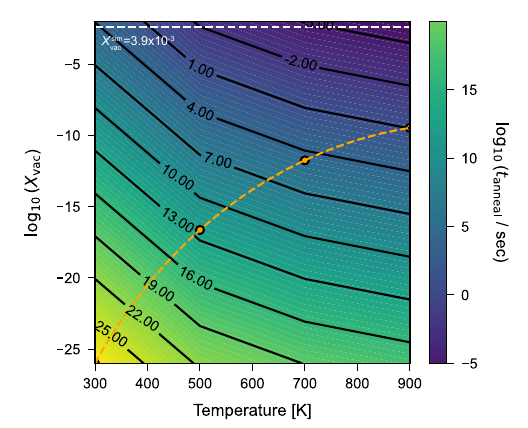}
\caption{
Temperature–vacancy concentration–time map.  
The annealing time ($t_{\rm anneal}$) required to reach the equilibrium short-range order, ${\rm SRO}^{\rm Eq}(T)$, is shown as a function of temperature and vacancy concentration.  
The horizontal dashed line indicates the fixed vacancy concentration used in our simulations, while the orange dashed line represents the equilibrium vacancy concentration $X_{\rm vac}^{\rm Eq}(T_{\rm anneal})$ at each annealing temperature.
}
    \label{fig:time_vac_solo}
\end{figure*}

\clearpage

\end{document}